\RequirePackage{fix-cm}
\documentclass[smallextended]{svjour3}       
\smartqed  
\usepackage{graphicx}
 \journalname{Materials Theory}

\usepackage{amsmath,amssymb}
\usepackage{bm}
\usepackage{graphicx}
\graphicspath{{../Figures/}}             
\usepackage{tikz}
\usetikzlibrary{arrows,decorations.pathmorphing,backgrounds,positioning,fit,shapes,patterns,fadings,calc,3d,er}
\usepackage[percent]{overpic}
\usepackage{color}
\usepackage{subfig}
\usepackage{url} 

\newcommand{\angstrom}{\textup{\AA}}

\def \pd{\partial}
\newcommand{\dV}[0]{\, \text{d}V}

\newcommand{\Cb}[0]{\bm{\mathcal{C}}}
\newcommand{\Db}[0]{\bm{\mathcal{D}}}

\begin{document}

\title{The Green tensor of Mindlin's anisotropic 
   first strain gradient elasticity
}


\author{Giacomo Po         \and
        Nikhil Chandra  Admal  \and  Markus Lazar
}


\institute{Giacomo Po \at
              Department of Mechanical and Aerospace Engineering, University of California Los Angeles, Los Angeles, CA 90095, USA\\
              Department of Mechanical and Aerospace Engineering, University of Miami, Coral Gables, FL 33146, USA\\
              \email{gpo@ucla.edu}             \\
           \and
          Nikhil Chandra  Admal \at
              Department of Materials Science and Engineering, University of California Los Angeles, Los Angeles, CA 90095, USA
                            \email{admal002@g.ucla.edu}             \\
             \and
              Markus Lazar \at
              Department of Physics,  
Darmstadt University of Technology,
Hochschulstr.~6, D-64289 Darmstadt, Germany
              \email{lazar@fkp.tu-darmstadt.de}           
}

\date{Received: date / Accepted: date}

\maketitle

\begin{abstract}
\textcolor{magenta}{We} derive the  Green tensor of Mindlin's anisotropic first strain gradient elasticity. 
The Green tensor is valid for arbitrary anisotropic materials, with up to 21 elastic constants and 171 gradient elastic constants in the  general case of triclinic media. In contrast to its classical counterpart, the Green tensor is non-singular at the origin, and it converges to the classical tensor a few characteristic lengths away from the origin. Therefore, the  Green  tensor of Mindlin's  first strain gradient elasticity can be regarded as a physical regularization of the classical anisotropic Green tensor. The isotropic Green tensor and other special cases are recovered as particular instances of the general anisotropic  result. The  Green tensor is implemented numerically and applied to the Kelvin problem with elastic constants determined from interatomic potentials.  Results are compared to molecular statics calculations carried out with the same potentials.
\end{abstract}

\section{Introduction\label{sec:Intro}}

Green functions are  objects of fundamental importance in field theories, since they represent the fundamental solution of linear inhomogeneous  partial differential equations (PDEs) from which any particular  solution can be obtained via convolution with the source term \cite{Green}. Moreover, Green functions are the basis of important numerical methods for boundary value problems, such as the boundary element method \cite{Becker92}, and they provide ``flexible" boundary conditions for atomistic simulations \cite{Trinkle2008}. In  the context of linear elasticity, the Green function is a tensor-valued function of rank two, also known as the Green tensor. When contracted with a concentrated force acting at the origin, the Green tensor yields the displacement field in an infinite elastic medium. \cite{Kelvin}  first derived the  closed-form expression of the classical Green tensor  for isotropic materials. For anisotropic materials, \cite{LR} and \cite{Synge} were able to derive the Green tensor in terms of an integral expression over the 
equatorial circle of the unit sphere in Fourier space. \cite{Barnett}  extended this  result to  the first two derivatives
of the Green tensor,  and showed that the line-integral representation is well suited for numerical integration \textcolor{magenta}{(see also \cite{Bacon,Teodosiu})}. 

The Green tensor and its derivatives are singular at the origin, ultimately as a consequence of the lack of intrinsic length scales in the classical theory of elasticity. The unphysical singularities in the  elastic fields derived from the Green tensor hinder their applicability in nano-mechanics, including the elastic theory of defects such as cracks, dislocations and inclusions \cite{Mura,Askes2011}. 
Generalized elastic field theories with intrinsic length scales have been proposed in the context of micro-continuum theories \cite{Eringen1999},
non-local theories \cite{Eringen2002}, and gradient  theories \cite{Kroner1963,Mindlin64,Mindlin68,Mindlin72,MiEs1968}. In particular, Mindlin's anisotropic strain gradient elasticity has received renewed attention as a tool to solve engineering problems at the micro- and nano-scales for realistic materials \cite{polizzotto2018anisotropy}. Only recently, the structure of the gradient-elastic tensor has been rationalized for different material symmetry classes \cite{Auffray13}, and its atomistic representation and ensuing determination from interatomic potentials has become available \cite{Admal16}. 

The number of independent strain gradient elastic moduli ranges from 5 for isotropic materials, to 171 in the general case of triclinic materials. While simple expressions of the Green tensor exist for the isotropic case \cite{Rogula73,LP18}, and for simplified anisotropic theories \cite{LP14,LP15}, the Green tensor of the fully anisotropic theory of Mindlin's strain gradient elasticity has remained so far a rather elusive object. 
\cite{Rogula73} provided an expression for the Green tensor in gradient elasticity of arbitrary order, which involves a sum of terms associated with the roots of a certain characteristic polynomial. However, such representation renders its numerical implementation rather impractical, and it conceals the mathematical structure of the Green tensor in relationship to its classical counterpart. 

The  objective of this paper is to derive a  simple representation of the Green tensor of Mindlin's anisotropic first strain gradient elasticity, whose integral kernel involves only matrix operations suitable for efficient numerical implementation. 
Following a brief summary of  Mindlin's anisotropic first strain gradient elasticity in section \ref{MAGEOS}, we derive the matrix representation of the Green tensor in section \ref{GTAHN}. It is shown that the Green tensor is non-singular at the origin, while its first gradient is finite but discontinuous at the origin.  The classical tail of the Green tensor, as well as its classical limit for vanishing gradient parameters are easily recovered from the non-singular expression. In section \ref{specialCases} we demonstrate that the Green tensor generalizes other expressions found in the literature. In section \ref{Kelvin} we consider the Kelvin problem and compare the prediction of the Green tensor to atomistic calculations.

\section{Mindlin's anisotropic gradient elasticity 
\label{MAGEOS}}
 
Let us consider  an infinite elastic body in three-dimensional space  and  assume that the  gradient of the displacement field $\bm u$ is additively decomposed into an elastic distortion tensor  $\bm\beta$ and an inelastic\footnote{
The inelastic distortion comprises plastic effects, and is typically an incompatible field. When the inelastic distortion is absent the elastic distortion is compatible.
} \textcolor{magenta}{eigen-distortion} tensor $\bm \beta^*$:
\begin{align}
\label{uIJ}
\partial_ju_i=\beta_{ij}+\beta^*_{ij}\, .
\end{align}
In the linearized theory of Mindlin's form-II first strain gradient elasticity \cite{Mindlin64,Mindlin68,MiEs1968,Mindlin72}, 
the strain energy density of an homogeneous and
centrosymmetric\footnote{Due to the centrosymmetry, there is no coupling
  between $e_{ij}$ and $\pd_m e_{kl}$.} material is given by 
\begin{align}
\label{W-an}
\mathcal{W}(\bm e,\bm\nabla\bm e)
=\frac{1}{2}\, \mathbb{C}_{ijkl}e_{ij}e_{kl}+\frac{1}{2}\,  \mathbb{D}_{ijmkln}\pd_m e_{ij} \pd_n e_{kl}\, .
\end{align}
The strain energy density \eqref{W-an} is a function of the infinitesimal elastic strain tensor
\begin{align}
e_{ij}=\frac{1}{2}\left(\beta_{ij}+\beta_{ji}\right)\, ,
\end{align} 
and of its gradient $e_{ij,m}$. The tensor  $\mathbb{C}$ is the standard rank-four tensor of elastic constants. By virtue of the symmetries
\begin{align}
\label{C}
\mathbb{C}_{ijkl}=\mathbb{C}_{jikl}=\mathbb{C}_{ijlk}=\mathbb{C}_{klij}\,,
\end{align}
it possesses up to 21 independent constants \textcolor{magenta}{with units of $\text{eV}/\angstrom^3$}. The tensor 
$\mathbb{D}$ is the rank-six tensor of strain gradient elastic constants, with symmetries
\begin{align}
\label{D}
\mathbb{D}_{ijmkln}=\mathbb{D}_{jimkln}=\mathbb{D}_{ijmlkn}=\mathbb{D}_{klnijm}\,.
\end{align}
 \textcolor{magenta}{It has units of $\text{eV}/\angstrom$}. In the general case  of triclinic materials the number of independent constants in the tensor $\mathbb{D}$
is equal to 171 \cite{Auffray13}.

The  quantities conjugate to the elastic strain tensor and its gradient are the Cauchy stress tensor $\bm\sigma$ and the double stress tensor $\bm\tau$, respectively. These are defined as:
\begin{align}
\label{CR1}
\sigma_{ij}&=\frac{\pd \mathcal{W}}{\pd e_{ij}}
=\mathbb{C}_{ijkl}e_{kl}\, , \\
\label{CR2}
\tau_{ijm}&=\frac{\pd \mathcal{W}}{\pd (\pd_m e_{ij})}
=\,\mathbb{D}_{ijmkln} e_{kl,n}\, .
\end{align}

In the presence of a body forces density $\bm b$, 
the  static Lagrangian density of the system becomes:
\begin{align}
\mathcal{L}=-\mathcal{W}-\mathcal{V}
=-\frac{1}{2}\left(\mathbb{C}_{ijkl}\beta_{ij}\beta_{kl}+\mathbb{D}_{ijmkln} \beta_{ij,m}  \beta_{kl,n}\right)+u_ib_i\, ,
\end{align}
where 
\begin{align}
\label{V}
\mathcal{V}=-u_i b_i
\end{align}
is the potential of the body force.
The condition of static equilibrium  is expressed by the Euler-Lagrange equation
\begin{align}
\label{EL-u}
\frac{\delta \mathcal{L}}{\delta u_i}=\frac{\pd \mathcal{L}}{\pd u_i}
-\pd_j\, \frac{\pd \mathcal{L}}{\pd (\pd_j u_i)}
+\pd_k \pd_j\, \frac{\pd \mathcal{L}}{\pd (\pd_k \pd_j u_i)}=0\,.
\end{align}
In terms of the Cauchy  and double stress  tensors, Eq.~\eqref{EL-u} takes the following form \cite{Mindlin64}:
\begin{align}
\label{MindlinBE}
\partial_j\big(\sigma_{ij}-\partial_m\tau_{ijm}\big)+b_i=0\, .
\end{align}
Using Eqs.~\eqref{uIJ} \eqref{CR1} \eqref{CR2}, Eq.~\eqref{MindlinBE} can be
cast in the following  equation for displacements:
\begin{align}
\label{u-LL}
L^{\text{M}}_{ik}\, u_k+f_i=0\,.
\end{align}
In Eq.~\eqref{u-LL}, $L^{\text{M}}_{ik}$ denotes the differential operator of Mindlin's anisotropic first strain gradient elasticity
\begin{align}
\label{LM}
L^{\text{M}}_{ik}=\mathbb{C}_{ijkl}\pd_j\pd_l-\mathbb{D}_{ijmkln}\pd_j\pd_l\pd_m\pd_n\, ,
\end{align}
while
\begin{align}
\label{effectiveF}
f_i=b_i-\left[\mathbb{C}_{ijkl}\partial_j-\mathbb{D}_{ijmkln}\partial_j\partial_m\partial_n\right]\beta^*_{kl}
\end{align}
is the forcing term. Note that the second term on the 
right hand side  of Eq.~\eqref{effectiveF} is an
``effective" internal force due to the inelastic eigen-distortion, 
and arises in the presence of material defects, such as inclusions, cracks, and dislocations. 
This term is the gradient version of  the internal force 
in Mura's eigen-strain \textcolor{magenta}{theory} \cite{Mura}.

\section{The Green tensor of Mindlin's first strain gradient elasticity
\label{GTAHN}} 
In this section, we derive the three-dimensional Green tensor of the operator~\eqref{LM}.
To this end,
we seek the solution to Eq.~\eqref{u-LL} in the form
\begin{align}
\label{convolutionU}
u_k=G_{kj}*f_j\, ,
\end{align}
where the symbol $*$ indicates convolution over the three-dimensional space, and
$\bm G$ is the Green tensor of Mindlin's anisotropic differential operator $\bm L^M$. Substituting
Eq.~\eqref{convolutionU} into Eq.~\eqref{u-LL}, 
one finds that $\bm G$ satisfies the following inhomogeneous  PDE:
\begin{align}
\label{G-LL}
\left[\mathbb{C}_{ijkl}\partial_j\partial_l-\mathbb{D}_{ijmkln}\partial_j\partial_l\partial_m\partial_n\right]G_{km}+\delta_{im} \delta=0\,.
\end{align}
In Eq.~\eqref{G-LL}, $\delta_{ij}$ is the Kronecker symbol, while $\delta$ is the three-dimensional Dirac $\delta$-distribution.

Taking the Fourier transform\footnote{
The  Fourier transform  and its inverse are defined as, respectively~\cite{Wl}:
\begin{align}
\hat{f}(\bm k)&=\int_{\mathbb{R}^3} f(\bm x)\, \text{e}^{-\text{i}\bm k\cdot\bm x}\dV\, ,\\
f(\bm x)&=\frac{1}{(2\pi)^3}\int_{\mathbb{R}^3} \hat{f}(\bm k)\, \text{e}^{\text{i}\bm k\cdot\bm x}\, \text{d}\hat{V}\,.
\end{align}
For a real-valued function, the inverse Fourier transform is
\begin{align}
f(\bm x)=\frac{1}{(2\pi)^3}\int_{\mathbb{R}^3} \hat{f}(\bm k)\, \cos\left(\bm k\cdot\bm x\right)\, \text{d}\hat{V}\,.
\end{align}
} of  Eq.~\eqref{G-LL}, we obtain \textcolor{magenta}{the following} algebraic equation for  the Green tensor $\hat{G}_{kj}(\bm k)$ in Fourier space
\begin{align}
\label{LN-FT}
\left[\mathcal{C}_{ik}(\bm k) + \mathcal{D}_{ik}(\bm k)\right]\hat{G}_{kj}(\bm k)=\delta_{ij}\,,
\end{align}
where 
\begin{align}
\mathcal{C}_{ik}(\bm k)
&=\mathbb{C}_{ijkl} k_j k_l\, ,\\
\mathcal{D}_{ik}(\bm k)
&=\mathbb{D}_{ijmkln} k_j k_lk_m k_n
\end{align}
are  symmetric matrices. 
If we further define the unit vector in Fourier space
\begin{align}
\bm \kappa=\frac{\bm k}{k}\,,\qquad k=\sqrt{k_ik_i}\, ,\qquad \bm\kappa^2=1 \,,
\end{align}
then \eqref{LN-FT} becomes:
\begin{align}
\label{LN-FT1}
k^2\left[
\mathcal{C}_{ik}(\bm \kappa) + k^2 \mathcal{D}_{ik}(\bm \kappa)\right]\hat{G}_{kj}(\bm k)=\delta_{ij}\,,
\end{align}
or equivalently, in matrix notation,
\begin{align}
k^2\left[
\Cb(\bm\kappa)
+k^2\Db(\bm\kappa)\right]\hat{\bm G}(\bm k)=\bm I\, .
\label{LN-FT11}
\end{align}
Stability of the differential operator $\bm L^M$ requires that the matrix 
$\Cb(\bm \kappa)+k^2\Db(\bm \kappa)$ be positive definite. Since this requirement must hold for  all $k$ and $\bm \kappa$, then the matrices $\Cb(\bm \kappa)$ and $\Db(\bm \kappa)$ must be individually positive definite. 
Under \textcolor{magenta}{the assumption that $\Cb(\bm \kappa)$ and $\Db(\bm \kappa)$ are symmetric positive definite (SPD) matrices}, the solution of \eqref{LN-FT11} in Fourier space  clearly reads:
\begin{align}
\hat{\bm G}(\bm k)=\frac{\left[
\Cb(\bm\kappa)+k^2
\Db(\bm\kappa)\right]^{-1}}{k^2}\, .
\label{LN-FT12}
\end{align}

The three-dimensional Green tensor in real space is  obtained 
by inverse Fourier transform of Eq.~\eqref{LN-FT12}. \textcolor{magenta}{It reads:}
\begin{align}
\bm G(\bm x)
&=\frac{1}{8\pi^3}\int_{\mathbb{R}^3} 
\frac{\left[
\Cb(\bm\kappa)+k^2
\Db(\bm\kappa)
\right]^{-1}}{k^2}\, \cos\left(\bm k\cdot\bm x\right)\, \text{d}\hat{V}\nonumber\\ 
&=\frac{1}{8\pi^3}\int_{\mathcal{S}}\int_0^\infty \left[
\Cb(\bm\kappa)+k^2
\Db(\bm\kappa)
\right]^{-1}\, \cos\left(k\bm \kappa\cdot\bm x\right)\, \text{d}k\, \text{d}{\omega}\, .
\label{Greal1}
\end{align}
In Eq.~\eqref{Greal1}, $\text{d}\hat{V}=k^2\, \text{d}k\, \text{d}\omega$ indicates the volume element in Fourier space,  and
$\text{d}\omega$ is  an elementary solid angle on the unit sphere
$\mathcal{S}$.  Our objective now is  to obtain an alternative expression of \textcolor{magenta}{the matrix inverse}
$[\Cb(\bm\kappa)+k^2
\Db(\bm\kappa)]^{-1}$ which allows us to carry out the the $k$-integral analytically. By doing so, the non-singular \textcolor{magenta}{nature} of the Green tensor at the origin is revealed.  We start by observing that, by virtue of its SPD character, the matrix  $\Cb(\bm\kappa)$  admits the following eigen-decomposition
\begin{align}
\Cb(\bm\kappa)=\bm R (\bm\kappa)\bm V^2(\bm\kappa)\bm R ^T(\bm\kappa)
\, ,
\end{align}
where $\bm R (\bm\kappa)$ is the orthogonal matrix of the eigenvectors of $\Cb(\bm\kappa)$, while $\bm V^2(\bm\kappa)$ is the diagonal matrix of positive eigenvalues of $\Cb(\bm \kappa)$. Moreover, the matrix
\begin{align}
\Cb^\frac{1}{2}=\bm R (\bm\kappa)\bm V(\bm\kappa)\bm R ^T(\bm\kappa)
\label{CHalf}
\end{align}
is also SPD.
Using \eqref{CHalf}, let us consider the following identity:
\begin{align}
\Cb+k^2\Db(\bm\kappa)=
\Cb^\frac{1}{2}
\left[\bm I+k^2\bm \Lambda^2 (\bm \kappa)\right]
\Cb^\frac{1}{2}\, ,
\end{align}
where
\begin{align}
\bm \Lambda^2 (\bm \kappa)=\Cb^{-\frac{1}{2}}(\bm\kappa)\Db(\bm\kappa)\Cb^{-\frac{1}{2}}(\bm\kappa)
\end{align}
is a SPD matrix with units of length squared. With this decomposition, the Green tensor in Fourier space becomes
\begin{align}
\hat{\bm G}(\bm k)
&=\Cb^{-\frac{1}{2}}(\bm \kappa)\frac{\left[\bm I+k^2\bm \Lambda^2(\bm \kappa)\right]^{-1}}{k^2}\Cb^{-\frac{1}{2}}(\bm \kappa)\, ,
\label{PLPT}
\end{align}
while in real space we obtain
\begin{align}
\bm G(\bm x)
=\frac{1}{8\pi^3}\int_{\mathcal{S}} 
\Cb^{-\frac{1}{2}}(\bm \kappa)
\int_0^\infty \left[\bm I+k^2\bm\Lambda^2(\bm\kappa)\right]^{-1}\, \cos(k\bm \kappa\cdot\bm x)\,  \text{d}k\, 
\Cb^{-\frac{1}{2}}(\bm \kappa)\text{d}\omega\, .
\label{NavierReal1}
\end{align}
In order to carry out the $k$-integral, we make use of the following matrix identity:\footnote{
The proof of \eqref{k-int} descends from the fact that 
$\bm  \Lambda^2 (\bm \kappa)$ is a real SPD matrix, and therefore it admits  the eigen-decomposition 
\begin{align}
\textcolor{magenta}
{\bm \Lambda^2 (\bm \kappa)={\bm Q}(\bm\kappa)\bm D^2(\bm\kappa){\bm Q}^{T}(\bm\kappa)\, ,}
\label{eigenDec}
\end{align}
where $\bm D^2(\bm\kappa)=\text{diag}\left\{\lambda^2_i(\bm\kappa)\right\}$ is the diagonal matrix of the positive eigenvalues of $\bm \Lambda^2 (\bm\kappa)$, and $\textcolor{magenta}{{\bm Q}(\bm\kappa)}$ is the orthogonal matrix of its eigenvectors. 
With this observation, we immediately obtain
\begin{align*}
\int_{0}^\infty 
\left[\bm I+k^2\bm\Lambda^2(\bm\kappa)\right]^{-1}\, \cos(k\bm \kappa\cdot\bm x)\, \text{d}k 
&=\int_{0}^\infty 
\left[\bm Q(\bm\kappa)\left(\bm I+k^2\bm D^2(\bm\kappa)\right)\bm Q^T(\bm\kappa)\right]^{-1}\, \cos(k\bm \kappa\cdot\bm x)\, \text{d}k\\
&=\bm Q(\bm\kappa)\int_{0}^\infty 
\text{diag}\left\{\frac{\cos(k\bm \kappa\cdot\bm x)}{1+k^2\lambda^2_i(\bm\kappa)}\right\}\, \text{d}k \, \bm Q^T(\bm\kappa) \, .
\end{align*}
\textcolor{magenta}{With the help of the definite integral 3.767 in \cite{GradshteynRyzhik}, we obtain}
\begin{align*}
\int_{0}^\infty 
\left[\bm I+k^2\bm\Lambda^2(\bm\kappa)\right]^{-1}\, \cos(k\bm \kappa\cdot\bm x)\, \text{d}k 
&=\frac{\pi}{2}\,\bm Q(\bm\kappa)\, \text{diag}\left\{\frac{\text{e}^{-|\bm \kappa\cdot\bm x|/\lambda_i(\bm \kappa)}}{\lambda_i(\bm \kappa)}\right\}\bm Q^T(\bm\kappa)\\
&=\frac{\pi}{2}\,\bm Q(\bm\kappa)\,\text{diag}\left\{\text{e}^{-|\bm \kappa\cdot\bm x|/\lambda_i(\bm \kappa)}\right\}\bm D^{-1}(\bm\kappa)\, \bm Q^T(\bm\kappa)\\
&=\frac{\pi}{2}\,\bm Q(\bm\kappa)\,\exp\left\{-|\bm \kappa\cdot\bm x|\,\bm D^{-1}(\bm\kappa)\right\}\, \bm Q^T(\bm\kappa) \bm\Lambda^{-1}(\bm\kappa)\\
&=\frac{\pi}{2}\,\exp\left\{-|\bm \kappa\cdot\bm x|\,\bm \Lambda^{-1}(\bm\kappa)\right\}\, \bm\Lambda^{-1}(\bm\kappa).
\end{align*}
In the last step we have used the property that the matrix exponential is an isotropic tensor-valued function of its argument.
}
\begin{align}
\label{k-int}
\int_{0}^\infty 
\left[\bm I+k^2\bm\Lambda^2(\bm\kappa)\right]^{-1}\, \cos(k\bm \kappa\cdot\bm x)\, \text{d}k
=\frac{\pi}{2}\,\exp\left(-|\bm \kappa\cdot\bm x|\,\bm \Lambda^{-1}(\bm\kappa)\right)\, \bm\Lambda^{-1}(\bm\kappa)\, .
\end{align}
With this identity, the Green tensor takes the form
\begin{align}
\bm G(\bm x)
=\frac{1}{16\pi^2}\int_{\mathcal{S}} 
\Cb^{-\frac{1}{2}}(\bm \kappa)
\exp\left\{-|\bm \kappa\cdot\bm x|\,\bm \Lambda^{-1}(\bm\kappa)\right\}\, \bm\Lambda^{-1}(\bm\kappa)\, 
\Cb^{-\frac{1}{2}}(\bm \kappa)\, \text{d}\omega\, .
\label{GT}
\end{align}

\begin{figure}[t]
\centering
\includegraphics[width=0.45\textwidth]{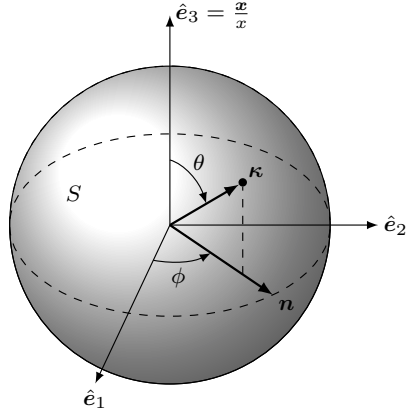}
\caption{The unit sphere in Fourier space. The unit vector $\bm \kappa(\theta,\phi)$ is defined by the azimuth angle $\phi$, and the  zenith angle $\theta$  measured from the axis $\hat{\bm e}_3=\bm x/x$.}
\label{kSpace}
\end{figure}

Next, Eq.~\eqref{GT} is further simplified noting that  the integration kernel  is an even \textcolor{magenta}{function} of $\bm\kappa$. Therefore, the integral over the unit sphere $\mathcal{S}$ is twice the  integral over a hemisphere. 

At the origin, any arbitrary  hemisphere $\mathcal{H}$  can be chosen, and the Green tensor assumes the value
\begin{align}
\bm G(\bm 0)
&=\frac{1}{8\pi^2}\int_{\mathcal{H}} 
\Cb^{-\frac{1}{2}}(\bm \kappa)
\, \bm\Lambda^{-1}(\bm\kappa)\, \Cb^{-\frac{1}{2}}(\bm \kappa)\text{d}\omega\, .
\end{align}
This noteworthy result shows that the Green tensor is
non-singular at the origin, \textcolor{magenta}{in contrast to classical elasticity}. 

Away from the origin, we can choose the  hemisphere having the direction $\bm x$ as the zenith.
This is a convenient choice  because all points \textcolor{magenta}{$\bm\kappa$ on such a} hemisphere satisfy \textcolor{magenta}{the condition} $\bm\kappa\cdot\bm x\ge0$. \textcolor{magenta}{This}  hemisphere \textcolor{magenta}{can be} parameterized by the  zenith angle $\theta$ and the azimuth angle $\phi$, as shown in Fig.~{\ref{kSpace}}. In this reference system,  \textcolor{magenta}{the unit vector $\bm \kappa$} can be expressed as
\begin{align}
\bm\kappa(\theta,\phi)=\sin\theta\cos\phi\, \hat{\bm e}_1+\sin\theta\sin\phi\, \hat{\bm e}_2+\cos\theta\, \hat{\bm e}_3\, ,
\end{align}
where $\hat{\bm e}_3=\bm x/x$. Finally, letting $q=\cos\theta$, the elementary solid angle becomes
\begin{align}
\text{d}\omega = \sin \theta\, \text{d}\theta\, \text{d}\phi =
-\text{d}q\, \text{d}\phi\, , 
\end{align}
and
\begin{align}
\bm\kappa(q,\phi)=\sqrt{1-q^2}\cos\phi\, \hat{\bm e}_1+\sqrt{1-q^2}\sin\phi\, \hat{\bm e}_2+q\, \hat{\bm e}_3\, .
\end{align}
Therefore the Green tensor of the anisotropic Mindlin differential operator of first order finally becomes
\begin{align}
\label{GT1}
{\bm G}(\bm x)
=\frac{1}{8\pi^2}\int_0^{2\pi}
\int_0^1 &
\Cb^{-\frac{1}{2}}(\bm \kappa)
\exp\left\{-qx\,\bm \Lambda^{-1}(\bm\kappa)\right\}\, \bm\Lambda^{-1}(\bm\kappa)\, 
\Cb^{-\frac{1}{2}}(\bm \kappa)
\, \text{d} q\, \text{d}\phi\,.
\end{align}

\subsection{The first two gradients of the Green tensor}
The first two gradients of the Green tensor are computed directly by differentiation of \eqref{GT}. 
The first gradient reads
 \begin{align}
\bm\nabla \bm G(\bm x)
=-\frac{1}{16\pi^2}\int_{\mathcal{S}} &
\Cb^{-\frac{1}{2}}(\bm \kappa)
\exp\left\{-|\bm \kappa\cdot\bm x|\,\bm \Lambda^{-1}(\bm\kappa)\right\}\, \bm\Lambda^{-2}(\bm\kappa)
\nonumber\\
&\times
\Cb^{-\frac{1}{2}}(\bm \kappa)\otimes \bm\kappa \, \text{sign}(\bm\kappa\cdot\bm x)\,\text{d}\omega\, .
\label{dGT}
\end{align}
In components this is:
 \begin{align}
 G_{ij,m}(\bm x)
=-\frac{1}{16\pi^2}\int_{\mathcal{S}} &\left[
\Cb^{-\frac{1}{2}}(\bm \kappa)
\exp\left\{-|\bm \kappa\cdot\bm x|\,\bm \Lambda^{-1}(\bm\kappa)\right\}\, \bm\Lambda^{-2}(\bm\kappa)\right.
\nonumber\\
&\times\left.\Cb^{-\frac{1}{2}}(\bm \kappa)\right]_{ij} \kappa_m \, \text{sign}(\bm\kappa\cdot\bm x)\,\text{d}\omega\, .
\label{dGT2}
\end{align}
Note that, because of the presence of the sign function, the gradient of the Green tensor is finite but discontinuous at the origin. 
From a computational perspective, it is more convenient to express this result in reference system of Fig.~\ref{kSpace}.   Doing so we find the alternative representation
 \begin{align}
 G_{ij,m}(\bm x)
=-\frac{1}{8\pi^2}\int_0^{2\pi}\int_0^1 &\left[
\Cb^{-\frac{1}{2}}(\bm \kappa)
\exp\left\{-|\bm \kappa\cdot\bm x|\,\bm \Lambda^{-1}(\bm\kappa)\right\}\, \bm\Lambda^{-2}(\bm\kappa)\right.
\nonumber\\
&\left.\Cb^{-\frac{1}{2}}(\bm \kappa)\right]_{ij} \kappa_m \,\text{d}q\,\,\text{d}\phi .
\label{dGT2}
\end{align}

 The second gradient \textcolor{magenta}{of the Green tensor} reads
 \begin{align}
\bm\nabla\bm\nabla \bm G(\bm x)
&=\frac{1}{16\pi^2}\int_{\mathcal{S}} \Big(
\Cb^{-\frac{1}{2}}(\bm \kappa)
\exp\left\{-|\bm \kappa\cdot\bm x|\,\bm \Lambda^{-1}(\bm\kappa)\right\}\, \nonumber\\
&\hspace{2cm}
\times
\bm\Lambda^{-3}(\bm\kappa)\,
\Cb^{-\frac{1}{2}}(\bm \kappa)\otimes \bm\kappa \otimes\bm\kappa
\nonumber\\
&\ 
-\Cb^{-\frac{1}{2}}(\bm \kappa)\,
 \bm\Lambda^{-2}(\bm\kappa)\, \Cb^{-\frac{1}{2}}(\bm \kappa)
\otimes \bm\kappa \otimes \bm\kappa 
\, \delta(\bm\kappa\cdot\bm x)
\Big)\,\text{d}\omega\, .
\label{ddGT}
\end{align}
\textcolor{magenta}{Letting $\bm n(\phi)=\bm \kappa(\pi/2,\phi)$ be a unit vector on the equatorial plane $\bm\kappa\cdot\bm x=0$, we finally obtain}
 \begin{align}
\bm\nabla\bm\nabla \bm G(\bm x)
&=\frac{1}{16\pi^2}\int_{\mathcal{S}} 
\Cb^{-\frac{1}{2}}(\bm \kappa)
\exp\left\{-|\bm \kappa\cdot\bm x|\,\bm \Lambda^{-1}(\bm\kappa)\right\}\, \nonumber\\
&\hspace{2cm}
\times
\bm\Lambda^{-3}(\bm\kappa)\,
\Cb^{-\frac{1}{2}}(\bm \kappa)\otimes \bm\kappa \otimes\bm\kappa \,\text{d}\omega\,
\nonumber\\
&\ 
-\frac{1}{8\pi^2 x}\int_{0}^{2\pi} 
\Cb^{-\frac{1}{2}}(\bm n)\,
 \bm\Lambda^{-2}(\bm n)\, \Cb^{-\frac{1}{2}}(\bm n)
\otimes \bm n \otimes \bm n \,\text{d}\phi\, .
\label{ddGT-2}
\end{align}
\textcolor{magenta}{Note that the second gradient of the Green tensor is singular at the origin.}

\subsection{The classical limit}

It is now shown that Green tensor \eqref{GT} converges to the classical Green tensor $\bm G^0$ \cite{LR,Synge} when the field point $\bm x$ is sufficiently far  from the origin compared to the characteristic  length scales, that is  when
\begin{align}
|\bm\kappa\cdot\bm x|/\lambda_i\gg 1,
\label{classicalLimitcondition}
\end{align}
where $\lambda_i$ is an eigenvalue of $\bm \Lambda$, and $i=1,2,3$.
This important property  guarantees that the non-singular Green tensor \eqref{GT1} regularizes the classical anisotropic Green tensor in the far field. Moreover, as a special case satisfying  condition \eqref{classicalLimitcondition}, the classical Green tensor $\bm G^0$   is also  recovered in the limit of vanishing tensor of strain gradient coefficients $\mathbb{D}$. 
The  classical Green tensor $\bm G^0$ is \textcolor{magenta}{readily} recovered if we consider the limit\footnote{
Using the eigen-decomposition \eqref{eigenDec}:
\begin{align*}
&\lim_{\||\bm\kappa\cdot\bm x|\, \bm\Lambda^{-1}\|\rightarrow\infty}\,
\exp\left\{-|\bm \kappa\cdot\bm x|\,\bm \Lambda^{-1}(\bm\kappa)\right\}\, \bm\Lambda^{-1}(\bm\kappa)=\nonumber\\
&\lim_{\||\bm\kappa\cdot\bm x|\, \bm D^{-1}\|\rightarrow\infty}\,
\bm Q(\bm\kappa)\exp\left\{-|\bm \kappa\cdot\bm x|\,\bm D^{-1}(\bm\kappa)\right\}\, \bm D^{-1}(\bm\kappa)\bm Q^T(\bm\kappa)=\nonumber\\
& \lim_{|\bm\kappa\cdot\bm x|/\lambda_i\rightarrow \infty} \bm Q(\bm\kappa)\,
\text{diag}\left\{\frac{\exp\left\{-|\bm \kappa\cdot\bm x|/\lambda_{i}(\bm\kappa)\right\}}{\lambda_i(\bm\kappa)}\right\} \,\bm Q^T(\bm\kappa)=\nonumber\\
&  \bm Q(\bm\kappa)\,\frac{2\bm I}{x}\, \delta(\bm \kappa\cdot\hat{\bm x} ) \,\bm Q^T(\bm\kappa)
=\frac{2\bm I}{x}\, 
\delta(\bm \kappa\cdot\hat{\bm x} )\, .
\end{align*}
}
\begin{align}
\lim_{\||\bm\kappa\cdot\bm x|\, \bm\Lambda^{-1}\|\rightarrow\infty}\,
\exp\left\{-|\bm \kappa\cdot\bm x|\,\bm \Lambda^{-1}(\bm\kappa)\right\}\, \bm\Lambda^{-1}(\bm\kappa)
=\frac{2\bm I}{x}\, 
\delta(\bm \kappa\cdot\hat{\bm x} )\, ,
\label{DiracLimit}
\end{align}
where $\hat{\bm x} =\bm x/x$ and $\bm I$ is the identity tensor. In fact, the substitution of \eqref{DiracLimit} into
 \eqref{GT} yields
\begin{align}
\bm G(\bm x)
\rightarrow \bm G^0(\bm x) 
&=\frac{1}{8\pi^2 x}\int_{\mathcal{S}} \Cb^{-1}(\bm \kappa)\, \delta(\bm \kappa\cdot\bm x ) \, \text{d}\omega
=\frac{1}{8\pi^2 x}\, \int_0^{2\pi} \Cb^{-1}(\bm n)\, \text{d} \phi\, .
\label{GTclassical}
\end{align}
\textcolor{magenta}{Here we used again the notation} $\bm n(\phi)=\bm \kappa(\pi/2,\phi)$ \textcolor{magenta}{to indicate}  a unit vector on the equatorial plane $\bm\kappa\cdot\bm x=0$.
 Note that the span of integration can be reduced to the range $0\le\phi\le \pi$ using the symmetry $\Cb^{-1}(\bm n)=\Cb^{-1}(-\bm n)$.

\section{Special cases\label{specialCases}}
In this section we show that the Green tensor \eqref{GT} generalizes other results obtained in the literature.
 
\subsection{The weakly non-local Green tensor $\mathbf{G}^\text{NL}$}
Lazar and Po \cite{LP15} have considered a simplified strain gradient elasticity theory under the assumption
\begin{align}
\mathbb{D}_{ijmkln}=\mathbb{C}_{ijkl} {L}_{mn}\, ,
\end{align}
a framework which was named Mindlin's strain gradient elasticity with weak non-locality because of its relation to non-local theories \cite{LazarNonLocal,LazarAgiasofitou}.  The Green tensor  \eqref{GT} recovers our previous result as a special case. In fact, under the previous assumption, we have
\begin{align}
\bm \Lambda(\bm\kappa)= \bm I\, \sqrt{\bm\kappa^T\bm L\bm \kappa}\, ,
\end{align}
and 
\begin{align}
\exp\left\{-|\bm \kappa\cdot\bm x|\,\bm \Lambda^{-1}(\bm\kappa)\right\}\, \bm\Lambda^{-1}(\bm\kappa)
=\bm I \frac{\exp\left(-\frac{|\bm\kappa\cdot\bm x|}{\sqrt{\bm\kappa^T\bm L\bm \kappa}}\right)}{\sqrt{\bm\kappa^T\bm L\bm \kappa}}\nonumber\, .
\end{align}
Therefore the Green tensor becomes
\begin{align}
\bm G^\text{NL}(\bm R)
&=\frac{1}{16\pi^2}\int_{\mathcal{S}} \Cb^{-1}(\bm \kappa) \frac{\exp\left(-\frac{|\bm\kappa\cdot\bm x|}{\sqrt{\bm\kappa^T\bm L\bm \kappa}}\right)}{\sqrt{\bm\kappa^T\bm L\bm \kappa}} \, \text{d}\omega\, ,
\label{GT-NL}
\end{align}
which is the expression given in \cite{LP15}. 

\subsection{The Green tensor of anisotropic gradient elasticity of Helmholtz type $\mathbf{G}^\text{H}$}
An even  simpler theory, named Mindlin's gradient elasticity of Helmholtz type, has been proposed by \cite{LP14}. The theory is characterized by only one gradient length scale parameter $\ell$, which renders the tensor $\bm L$ diagonal:
\begin{align}
\bm L=\ell^2\, \bm I\, .
\label{LH}
\end{align}
The non-singular Green tensor of this theory is obtained by substituting \eqref{LH} in \eqref{GT-NL}, thus yielding
\begin{align}
\bm G^\text{H}(\bm R)
&=\frac{1}{16\pi^2 \ell}\int_{\mathcal{S}} \Cb^{-1}(\bm \kappa) \exp\left(-\frac{|\bm\kappa\cdot\bm x|}{\ell}\right) \, \text{d}\omega\, ,
\label{GT-H}
\end{align}
which coincides with the expression given in \cite{LP14}.

\subsection{The isotropic Green tensor $\mathbf{G}^\text{I}$}
\begin{figure}[t!]
\centering
\includegraphics[width=0.48\textwidth]{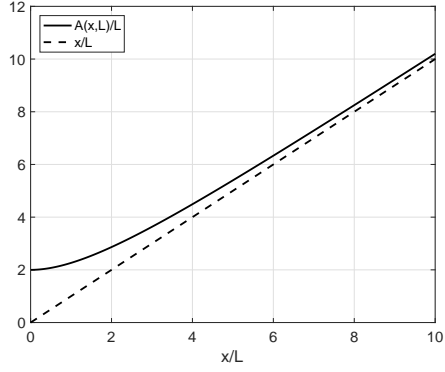}
\caption{Plot of the regularized distance function $A(x,\ell)$.}
\label{Afunc}
\end{figure}

The isotropic tensor $\mathbb{C}$ has components
\begin{align}
\label{C-iso}
\mathbb{C}_{ijkl}=\lambda\delta_{ij}\delta_{kl}
+\mu\big(\delta_{ik}\delta_{jl}+\delta_{il}\delta_{jk}\big)\,,
\end{align}
where $\lambda$ and $\mu$ are the Lam\'e constants. 
\textcolor{magenta}{On the other hand,}
the isotropic tensor $\mathbb{D}$ reads
\begin{align}
\label{D-iso}
\mathbb{D}_{ijmkln}&=
\frac{a_1}{2}\big(\delta_{ij}\delta_{km}\delta_{ln}
+\delta_{ij}\delta_{kn}\delta_{lm}
+\delta_{kl}\delta_{im}\delta_{jn}
+\delta_{kl}\delta_{in}\delta_{jm}\big)
\nonumber\\
&\
+\frac{a_3}{2}\big(\delta_{jk}\delta_{im}\delta_{kl}
+\delta_{ik}\delta_{jm}\delta_{nl}
+\delta_{il}\delta_{jm}\delta_{kn}
+\delta_{jl}\delta_{im}\delta_{kn}\big)
\nonumber\\
&\
+\frac{a_5}{2}\big(\delta_{jk}\delta_{in}\delta_{lm}
+\delta_{ik}\delta_{jn}\delta_{lm}
+\delta_{jl}\delta_{km}\delta_{in}
+\delta_{il}\delta_{km}\delta_{jn}\big)
\nonumber\\
&\ 
+2a_2\, \delta_{ij}\delta_{kl}\delta_{mn}
+a_4\big(\delta_{il}\delta_{jk}\delta_{mn}
+\delta_{ik}\delta_{jl}\delta_{mn}\big)\,,
\end{align}
where $a_1$, $a_2$, $a_3$, $a_4$, $a_5$ are the gradient 
parameters in isotropic Mindlin's first strain gradient elasticity theory
\cite{Mindlin64} (see also~\cite{Mindlin68,Lazar16}). 
Therefore, the matrices $\Cb(\bm\kappa)$ and $\Db(\bm\kappa)$ become, respectively
\begin{align}
\mathcal{C}_{ik}(\bm\kappa)
&=(\lambda+2\mu)\kappa_i\kappa_k
+\mu\big(\delta_{ik}-\kappa_i\kappa_k\big)\label{CFiso}\, ,\\
\mathcal{D}_{ik}(\bm\kappa)&=
2(a_1+a_2+a_3+a_4+a_5)\kappa_{i}\kappa_k
\nonumber
\\
&\ 
+\frac{1}{2}\,(a_3+2a_4+a_5)\big(\delta_{ik}-\kappa_{i}\kappa_k\big)
\nonumber\\
&=(\lambda+2\mu)\, \ell_1^2 \kappa_{i}\kappa_k
+\mu\, \ell_2^2 \big(\delta_{ik}-\kappa_{i}\kappa_k\big)\, .
\label{DFiso}
\end{align}
  \textcolor{magenta}{The two characteristic lengths $\ell_1$ and $\ell_2$ introduced above are defined as}
\begin{align}
\label{l1}
\ell_1^2&=\frac{2(a_1+a_2+a_3+a_4+a_5)}{\lambda+2\mu}\,,\\
\label{l2}
\ell_2^2&=\frac{a_3+2a_4+a_5}{2\mu}\,. 
\end{align}
Owing to the special structure\footnote{\label{footNoteSpecialStructure}
Consider a matrix $\bm A$ with structure 
\begin{align}
A_{ij}=a\kappa_i\kappa_j+b(\delta_{ij}-\kappa_i\kappa_j)\, .
\end{align}
If $a>b>0$, then the matrix is SPD, and a unique SPD square root of  $A_{ij}$ exists with form
\begin{align}
A_{ij}^{\frac{1}{2}}&=\sqrt{a}\kappa_i\kappa_j+\sqrt{b}(\delta_{ij}+\kappa_i\kappa_j)\, .
\end{align}
Moreover, the inverse of $A_{ij}$ reads
\begin{align}
A_{ij}^{-1}&=\frac{1}{a}\kappa_i\kappa_j+\frac{1}{b}(\delta_{ij}-\kappa_i\kappa_j)\, .
\end{align}
}
 of $\Cb(\bm\kappa)$ and $\Db(\bm\kappa)$,  the following results are easily obtained: 
\begin{align}
\mathcal{C}^{-\frac{1}{2}}_{ij}(\bm\kappa)
&=\frac{1}{\sqrt{\mu}}\left(\delta_{ij}-\kappa_i\kappa_j\right)-\frac{1}{\sqrt{\lambda+2\mu}}\kappa_i\kappa_j\\
%
\Lambda_{ij}^{-1}(\bm\kappa)&=\frac{1}{\ell_2}\left(\delta_{ij}-\kappa_i\kappa_j\right)+\frac{1}{\ell_1}\kappa_i\kappa_j\, .
\end{align}
The matrix $\bm\Lambda^{-1}$ admits the eigenvalue $1/\ell_1$, corresponding to the eigenvector $\hat{\bm v}_1=\bm\kappa$. The degenerate eigenvalue $1/\ell_2$ has multiplicity two, corresponding to two arbitrary eigenvectors $\hat{\bm v}_2$ and $\hat{\bm v}_3$ perpendicular to $\bm \kappa$. Choosing such eigenvectors to be mutually orthogonal, the matrix  
 $\bm \Lambda^{-1}$ admits the eigen decomposition $\bm \Lambda^{-1}=\bm Q \bm D^{-1} \bm Q^T$. Here 
\begin{align}
\bm Q=[\hat{\bm v}_1\, \hat{\bm v}_2\, \hat{\bm v}_3]
\end{align}
is an orthogonal  matrix whose columns are the  eigenvectors of $\bm \Lambda^{-1}$, and
\begin{align}
\bm D^{-1}=\text{diag}\left\{\frac{1}{\ell_1},\, \frac{1}{\ell_2},\,\frac{1}{\ell_2}\right\}
\end{align}
is the diagonal matrix of its eigenvalues. This special form of $\bm Q$ yields the identity
\begin{align}
\Cb^{-\frac{1}{2}}\bm Q=\bm Q\, \text{diag}\left\{-\frac{1}{\sqrt{\lambda+2\mu}},\, \frac{1}{\sqrt{\mu}},\,\frac{1}{\sqrt{\mu}}\right\}\, .
\end{align}
 Using these results in \eqref{GT}, we obtain
\begin{align}
\bm G^I(\bm x)
&=\frac{1}{16\pi^2}\int_{\mathcal{S}} 
\Cb^{-\frac{1}{2}}\bm Q\exp\left\{-|\bm \kappa\cdot\bm x|\,\bm D^{-1}\right\}\, \bm D^{-1}\bm Q^T
\Cb^{-\frac{1}{2}}\text{d}\omega\nonumber\\
&=\frac{1}{16\pi^2}\int_{\mathcal{S}} 
\bm Q\, \text{diag}\left\{\frac{e^\frac{-|\bm \kappa\cdot\bm x|}{\ell_1}}{\ell_1(\lambda+2\mu)},\, \frac{e^\frac{-|\bm \kappa\cdot\bm x|}{\ell_2}}{\ell_2\mu},\,\frac{e^\frac{-|\bm \kappa\cdot\bm x|}{\ell_2}}{\ell_2\mu}\right\}
\bm Q^T\text{d}\omega\nonumber\\
&=\frac{1}{16\pi^2}\int_{\mathcal{S}} \frac{e^\frac{-|\bm \kappa\cdot\bm x|}{\ell_1}}{(\lambda+2\mu)\ell_1}\hat{\bm v}_1\otimes\hat{\bm v}_1\text{d}\omega\nonumber\\
&+\frac{1}{16\pi^2}\int_{\mathcal{S}} \frac{e^\frac{-|\bm \kappa\cdot\bm x|}{\ell_2}}{\mu\ell_2}\left(\hat{\bm v}_2\otimes\hat{\bm v}_2+\hat{\bm v}_3\otimes\hat{\bm v}_3\right)\text{d}\omega\, .
\label{GTI}
\end{align}
Because they \textcolor{magenta}{form} an orthonormal basis, the three eigenvectors satisfy \textcolor{magenta}{the identity}  $\hat{\bm v}_1\otimes\hat{\bm v}_1+\hat{\bm v}_2\otimes\hat{\bm v}_2+\hat{\bm v}_3\otimes\hat{\bm v}_3=\bm I$, hence we have
\begin{align}
\bm G^I(\bm x)
&=\frac{1}{16\pi^2}\int_{\mathcal{S}} \left[\frac{e^\frac{-|\bm \kappa\cdot\bm x|}{\ell_1}}{(\lambda+2\mu)\ell_1}{\bm \kappa}\otimes{\bm \kappa}+\frac{e^\frac{-|\bm \kappa\cdot\bm x|}{\ell_2}}{\mu\ell_2}\left(\bm I-{\bm \kappa}\otimes{\bm \kappa}\right)\right]
\text{d}\omega\, .
\label{GTI2}
\end{align}
The integral over the unit sphere is carried out using the relation
\begin{align}
\int_{\mathcal{S}} \frac{e^\frac{-|\bm \kappa\cdot\bm x|}{\ell}}{\ell}\kappa_i\kappa_j\, \text{d}\omega
=2\pi\, \partial_i\partial_j A(x,\ell)\, ,
\label{KKintegral}
\end{align}
where the  scalar  function $A(x,\ell)$ is
\begin{align}
A(x,\ell)=x+\frac{2\ell^2}{x} -\frac{2\ell^2}{x}e^{-x/\ell}\, .
\label{Afunction}
\end{align}
The scalar function $A(x,\ell)$ can be regarded as a regularized distance function in the sense that $A(x,\ell)$ tends to $x$ when $x/\ell\gg 1$, while it smoothly approaches to $2\ell$ for small  $x$, as shown in Fig.~\ref{Afunc}.
By sake of \eqref{KKintegral}, the Green tensor finally becomes:
\begin{align}
\label{GT-iso}
G_{ij}(\bm x)=
\frac{1}{8\pi\mu}\, 
\Big[
\frac{\mu}{\lambda+2\mu}\, \pd_i\pd_j A(x,\ell_1)
+\big(\delta_{ij}\Delta-\pd_i\pd_j\big) A(x,\ell_2)\Big]\, .
\end{align}
This result can also be obtained by direct inverse Fourier transform of \eqref{LN-FT12}, as shown in Appendix \ref{DirectDerivationGTI}.
A \textcolor{magenta}{more} detailed analysis of the isotropic Green tensor \eqref{GT-iso} can be found in \cite{LP18}.

\section{A comparison with Molecular Statics: The  Kelvin problem \label{Kelvin}}

\newcommand{\ex}[1]{$\cdot 10^{#1}$}

\begin{table}[b!]
\centering
\begin{tabular}{| l | l | l |l |}
\hline
& Cu EAM  & Cu MEAM  & Al MEAM\\
\hline
${C}_{1,1}$ $[\text{eV}/\angstrom^3]$&1.0868  & 1.0994 &  7.1366\ex{-1}\\
${C}_{1,2}$  $[\text{eV}/\angstrom^3]$&  7.9386\ex{-1}&   7.7973\ex{-1} &  3.8649\ex{-1}\\
${C}_{4,4}$  $[\text{eV}/\angstrom^3]$ &5.2252\ex{-1}  & 5.1043\ex{-1}  & 1.9704\ex{-1}\\
\hline
${D}_{1,1}$    $[\text{eV}/\angstrom]$&1.1182   &6.5018\ex{-1}   &1.0855\\
${D}_{1,2}$     $[\text{eV}/\angstrom]$&  3.5814\ex{-1} &  3.6659\ex{-1} &  1.4572\ex{-1}\\
${D}_{1,3}$     $[\text{eV}/\angstrom]$&  3.7951\ex{-1} &  2.4150\ex{-1} &  1.5934\ex{-1}\\
${D}_{2,2}$     $[\text{eV}/\angstrom]$&  4.7935\ex{-1}  & 7.3885\ex{-1} &  8.4221\ex{-1}\\
${D}_{2,3}$   $[\text{eV}/\angstrom]$ &   3.0103\ex{-1}  & 2.0651\ex{-1} &  1.5671\ex{-1}\\
${D}_{2,4}$    $[\text{eV}/\angstrom]$&   1.2789\ex{-1} &  4.7496\ex{-1} &  7.1708\ex{-1}\\
${D}_{2,5}$    $[\text{eV}/\angstrom]$&   1.0652\ex{-1} & -4.2545\ex{-2} & -1.1434\ex{-2}\\
${D}_{3,3}$    $[\text{eV}/\angstrom]$&   4.3662\ex{-1} &  2.9055\ex{-1} &  2.7613\ex{-1}\\
${D}_{3,5}$     $[\text{eV}/\angstrom]$&  1.2789\ex{-1} & -1.8275\ex{-2} & -1.2408\ex{-1}\\
${D}_{16,16}$ $[\text{eV}/\angstrom]$&  1.4925\ex{-1} &  3.7419\ex{-2} &  1.6786\ex{-1}\\
${D}_{16,17}$ $[\text{eV}/\angstrom]$&  1.0652\ex{-1} &  3.7394\ex{-2} &  1.5006\ex{-1}\\
\hline
\end{tabular}
\caption{Elastic and gradient-elastic constants obtained from the interatomic potentials \cite{kimlee2001} and \cite{kimmendelev2008}. }
\label{ElasticTable}
\end{table}

In this section, we compare the Green tensor obtained from Mindlin's
strain gradient elastic theory to that obtained from an atomistic system.
This study was carried out using Minimol \cite{TadmorBook} which is a
KIM-compliant \textcolor{magenta}{molecular dynamics (MD) and molecular statics (MS)} program. The Open Knowledgebase of Interatomic Models (KIM) is a project focused on creating standards for atomistic simulations including an application programming interface (API) for information exchange between atomistic simulation codes and interatomic potentials \cite{Tadmor2011,TadmorKIM2013}.

\begin{figure*}[t!]
\centering
\subfloat[$\mathbb{C}$ for Cu EAM]{
\begin{tikzpicture}[scale=1]
\node[]  at (0,0) {\includegraphics[width=0.29\textwidth]{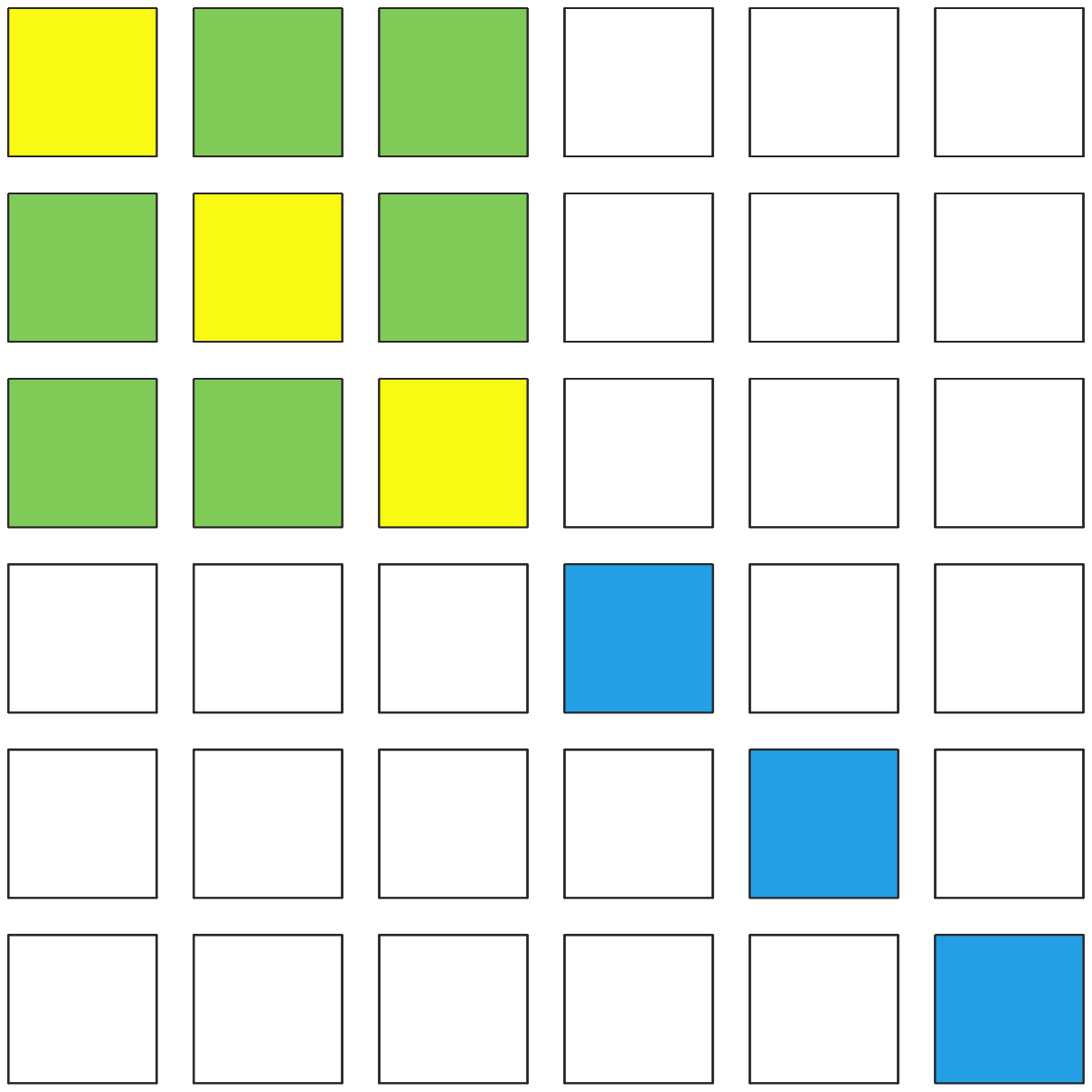}};
\end{tikzpicture}
\label{C_Cu_fcc_EAM_Mendelev_2013}
}
\subfloat[$\mathbb{C}$ for Cu MEAM]{
\begin{tikzpicture}[scale=1]
\node[]  at (0,0) {\includegraphics[width=0.29\textwidth]{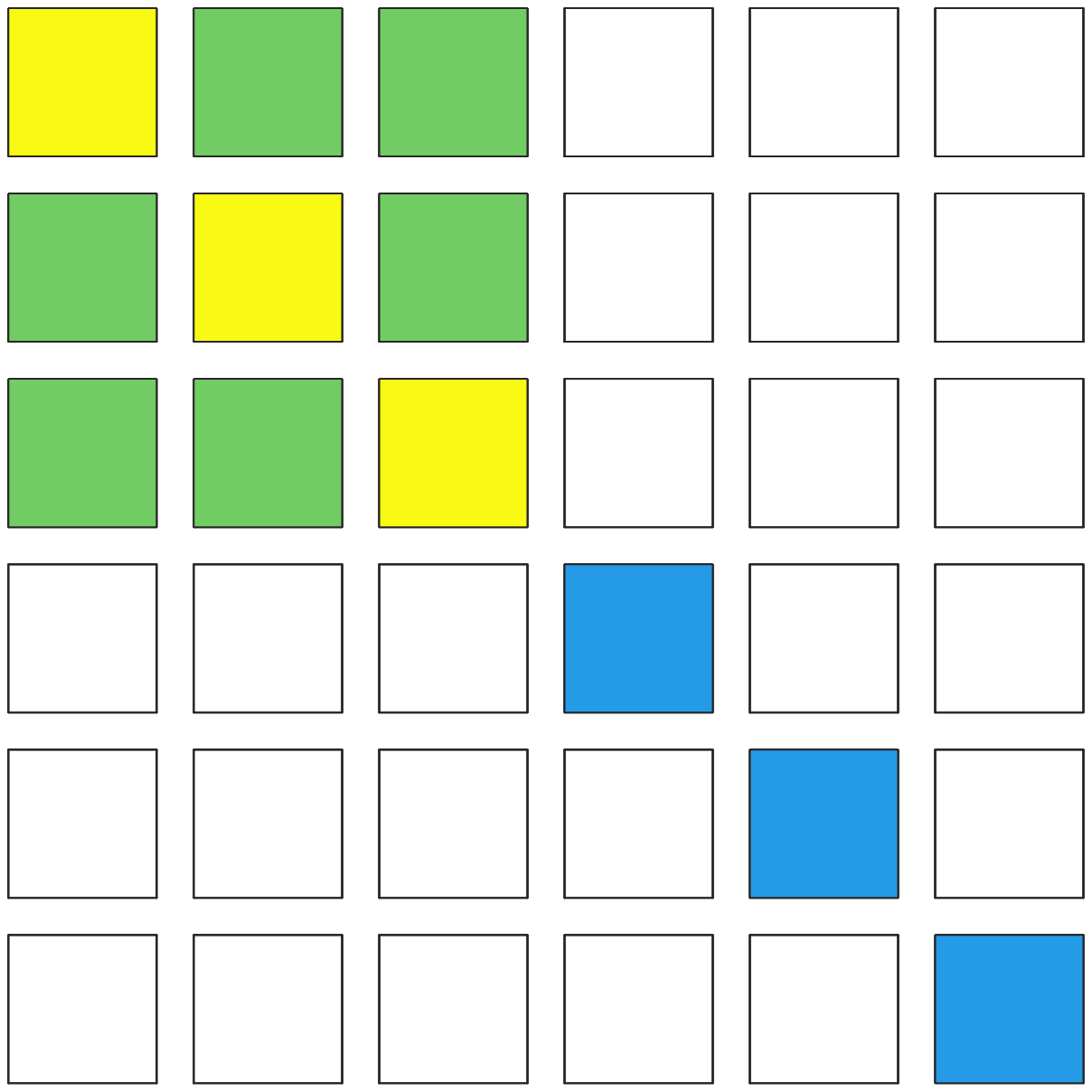}};
\end{tikzpicture}
\label{C_Cu_fcc_MEAM_Lee_2001}
}
\subfloat[$\mathbb{C}$ for Al MEAM]{
\begin{tikzpicture}[scale=1]
\node[]  at (0,0) {\includegraphics[width=0.33\textwidth]{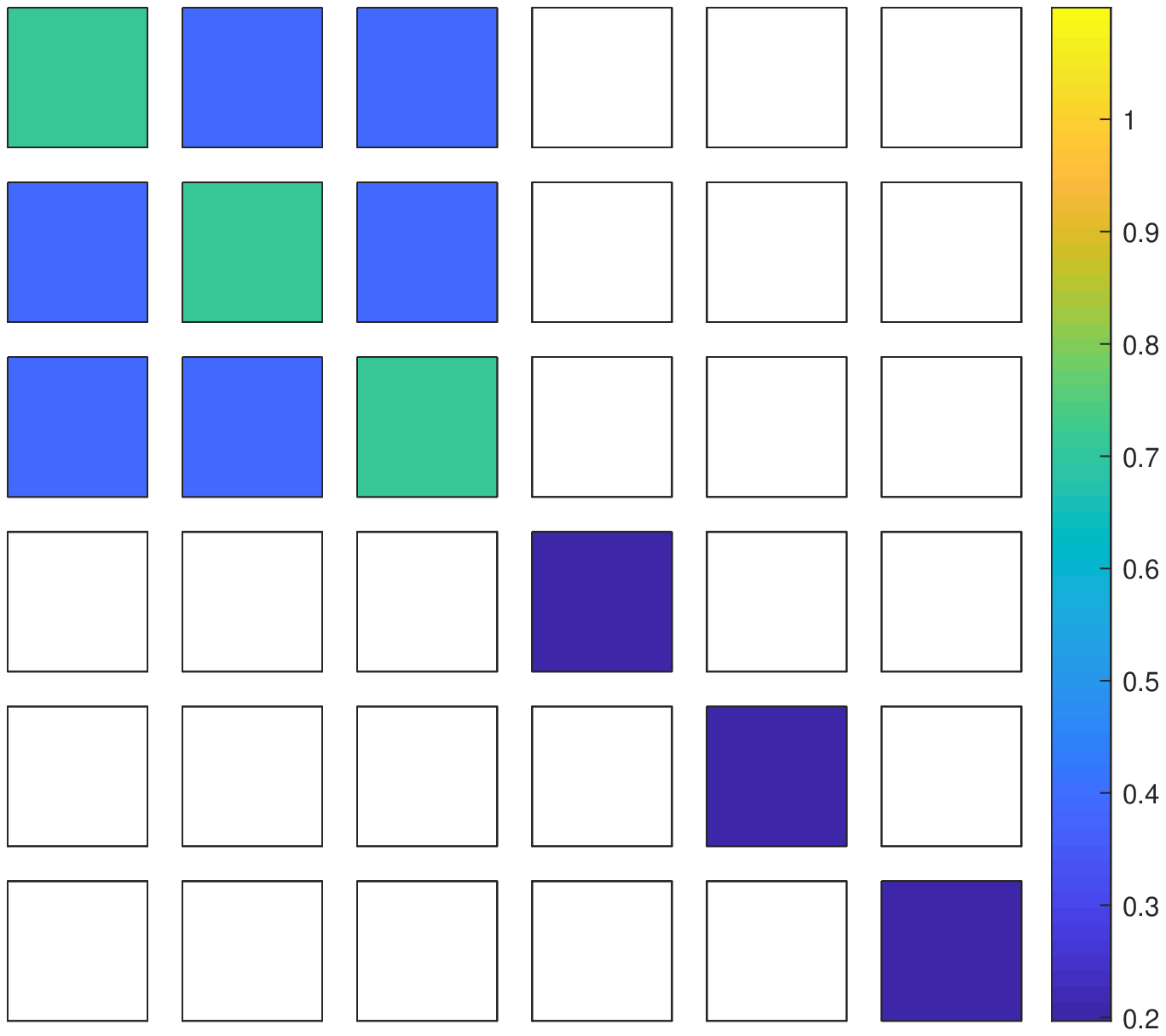}};
\node[]  at (2,1.8) {$\frac{\text{eV}}{\angstrom^3}$};
\end{tikzpicture}
\label{C_Al_fcc_MEAM_Lee_2001}
}\hfill
\subfloat[$\mathbb{D}$ for Cu EAM]{
\begin{tikzpicture}[scale=1]
\node[]  at (0,0) {\includegraphics[width=0.29\textwidth]{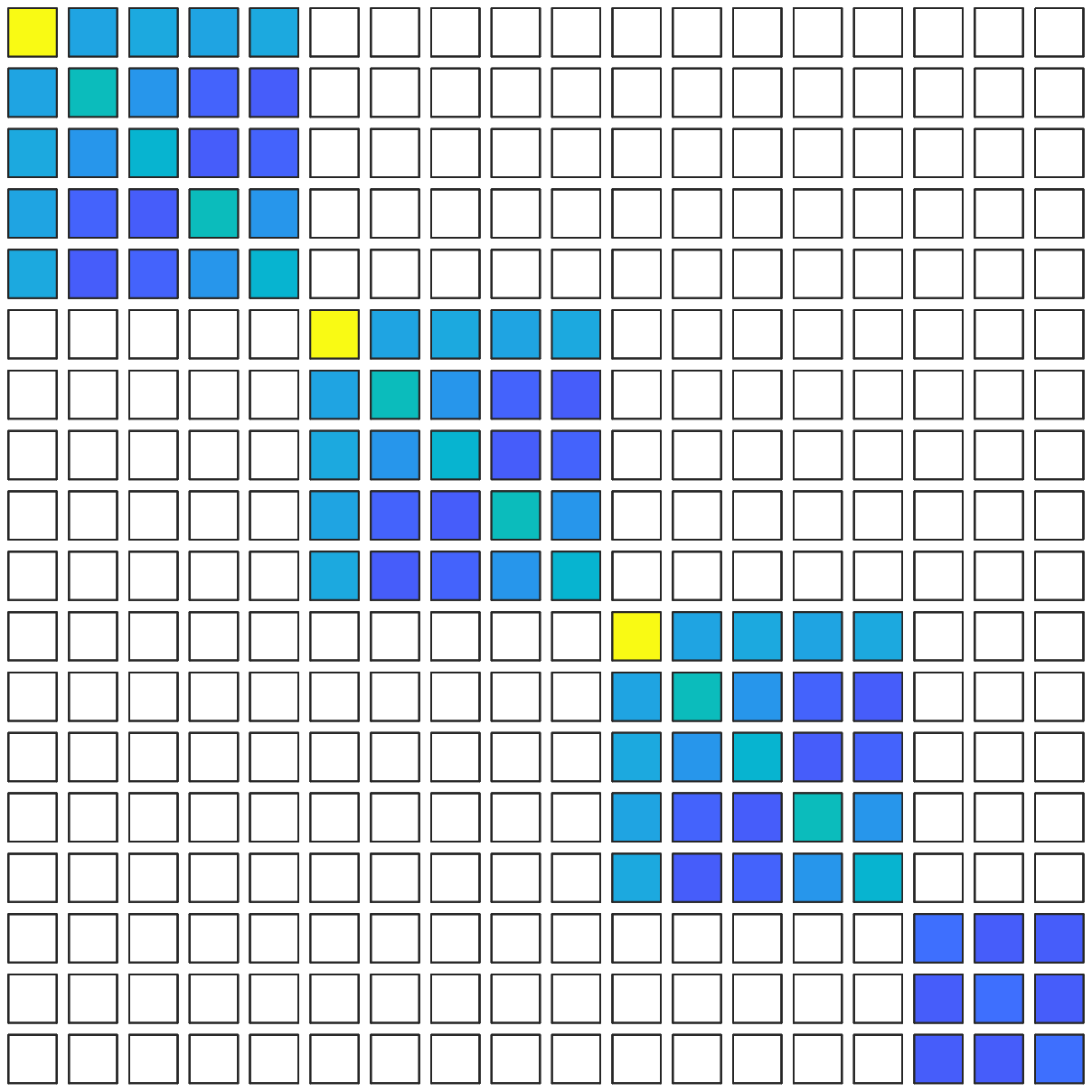}};
\end{tikzpicture}
\label{D_Cu_fcc_EAM_Mendelev_2013}
}
\subfloat[$\mathbb{D}$ for Cu MEAM]{
\begin{tikzpicture}[scale=1]
\node[]  at (0,0) {\includegraphics[width=0.29\textwidth]{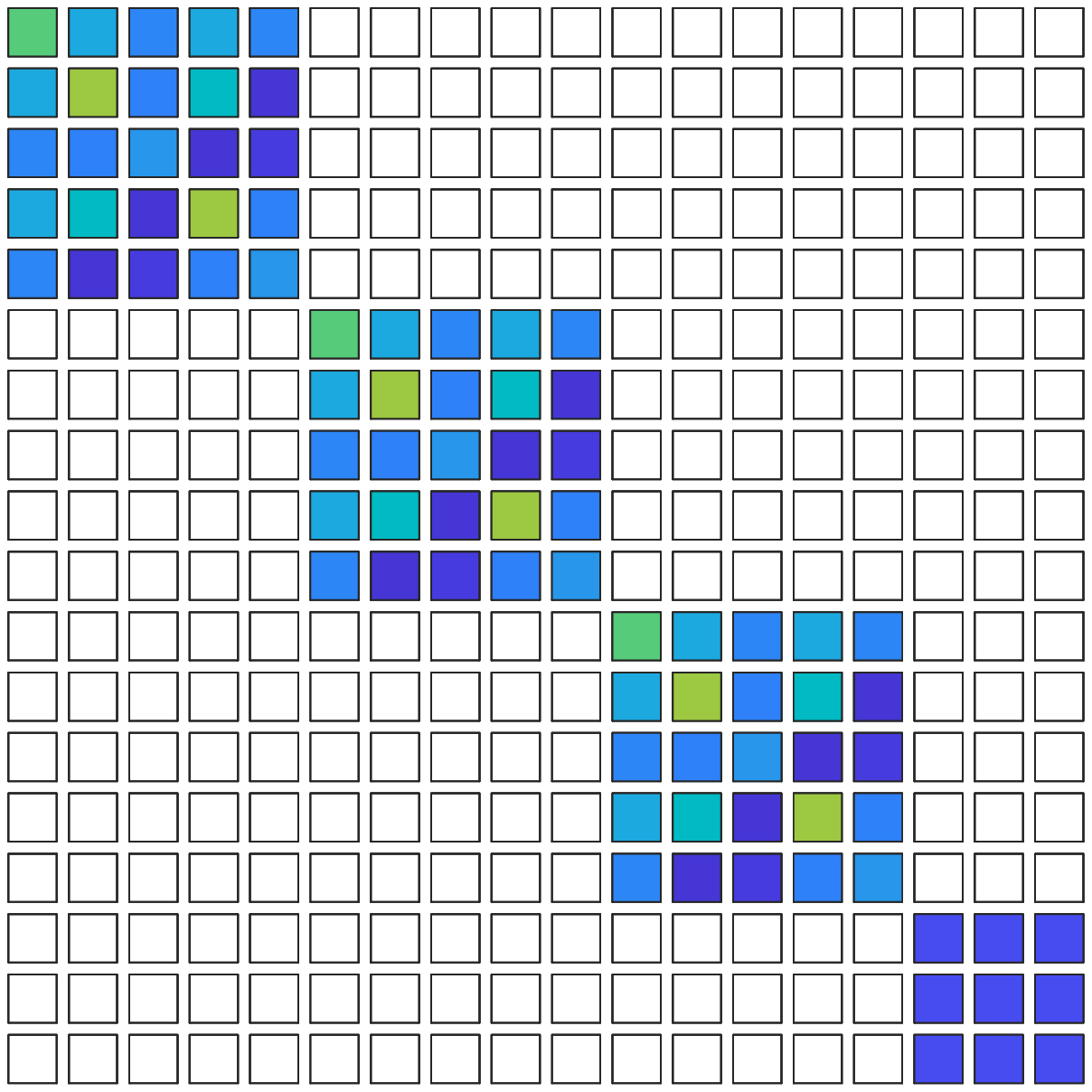}};
\end{tikzpicture}
\label{D_Cu_fcc_MEAM_Lee_2001}
}
\subfloat[$\mathbb{D}$ for Al MEAM]{
\begin{tikzpicture}[scale=1]
\node[]  at (0,0) {\includegraphics[width=0.33\textwidth]{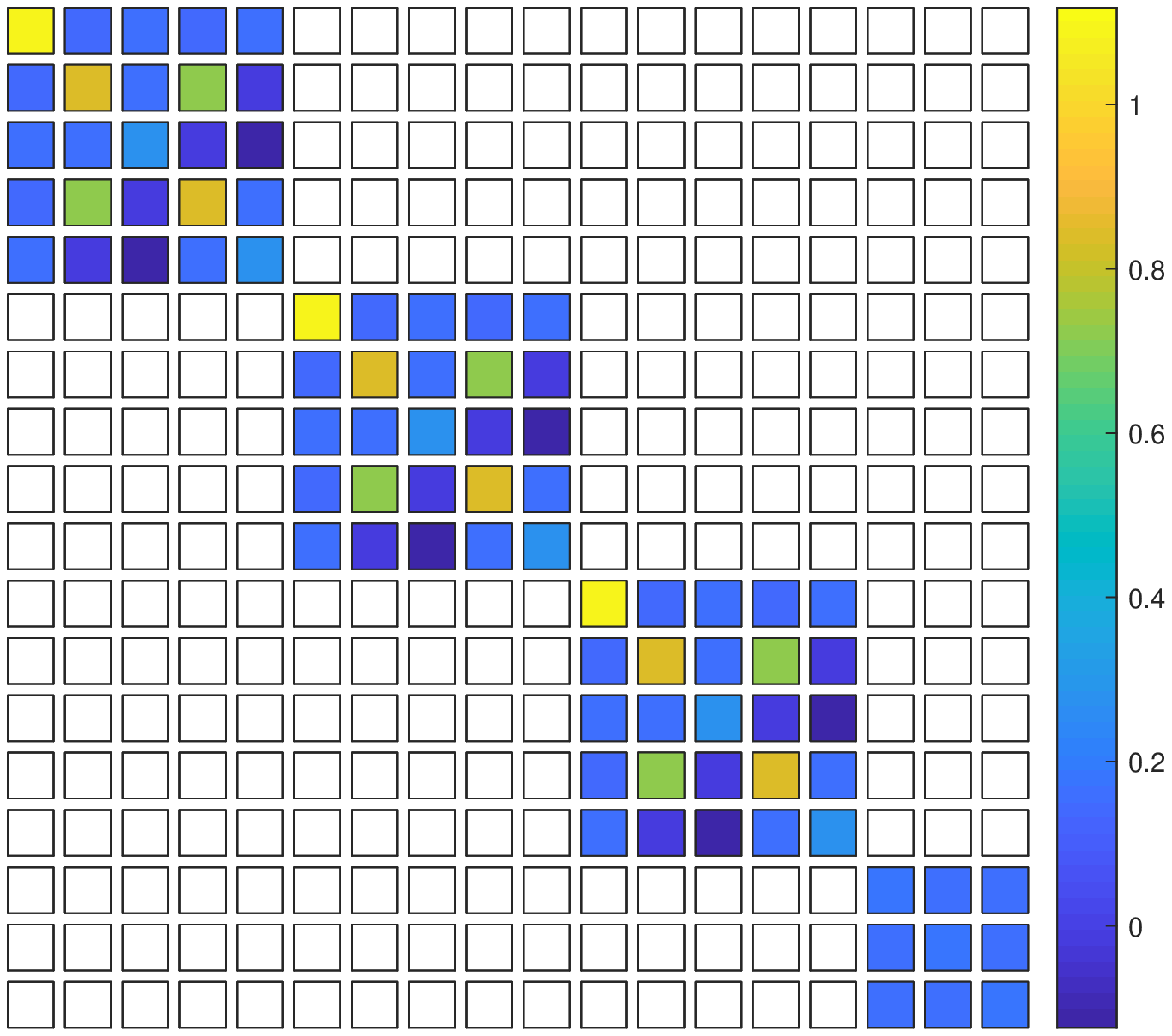}};
\node[]  at (2,1.8) {$\frac{\text{eV}}{\angstrom}$};
\end{tikzpicture}
\label{D_Al_fcc_MEAM_Lee_2001}
}\hfill
\caption{Voigt representation of the elastic tensors $\mathbb{C}$ and gradient-elastic tensor $\mathbb{D}$ for fcc Al and Cu, computed from the interatomic potentials \cite{kimlee2001} and \cite{kimmendelev2008}. 
\protect\subref{C_Cu_fcc_EAM_Mendelev_2013} and \protect\subref{D_Cu_fcc_EAM_Mendelev_2013} Cu for EAM potential \cite{kimmendelev2008}.
\protect\subref{C_Cu_fcc_MEAM_Lee_2001} and \protect\subref{D_Cu_fcc_MEAM_Lee_2001} Cu for MEAM potential \cite{kimlee2001}.
\protect\subref{C_Al_fcc_MEAM_Lee_2001} and \protect\subref{D_Al_fcc_MEAM_Lee_2001} Al for MEAM potential \cite{kimlee2001}.
}
\label{voigtCD}
\end{figure*}

We choose face-centered-cubic Aluminum and Copper for this comparison, and consider the
following two interatomic potentials: \textcolor{magenta}{the} modified-embedded-atom-method (MEAM) \textcolor{magenta}{by}
\cite{lee2001}, and the embedded-atom-potential  \textcolor{magenta}{by} \cite{mendelev2008}, \textcolor{magenta}{which are}
archived in the OpenKIM repository. Elastic and gradient-elastic constants for these potentials were computed using the method described in \cite{Admal16}, and they are available on the KIM repository \cite{kimlee2001,kimmendelev2008}. 
For convenience, the values of the independent elastic and gradient-elastic constants are reported in table \ref{ElasticTable}. These  components are used to populate the elastic tensors $\mathbb{C}$ and $\mathbb{D}$ \cite{Admal16,Auffray13}. The Voigt structure of the resulting tensors $\mathbb{C}$ and $\mathbb{D}$ is shown in Fig.~\ref{voigtCD}.

The atomistic
system is constructed by stacking together $15\times15\times15$ unit cells
resulting in $13500$ atoms.  A force of $0.0116$ eV/$\angstrom$ in the $x_1$ direction is imposed on the central atom of
the system, and displacement boundary conditions are imposed on five 
layers of atoms close to the boundary using the classical solution given in Eq.~\eqref{GTclassical}. 
The padding atoms thickness is 0.15 times the size of the box. 
A MS simulation is carried out using the above-mentioned boundary
conditions resulting in a deformed crystal. The resulting displacement field
normalized with respect to the force on the central atom yields the
atomistic Green tensor component fields.

\begin{figure*}[t!]
\centering
\subfloat[Cu EAM]{
\includegraphics[width=0.45\textwidth]{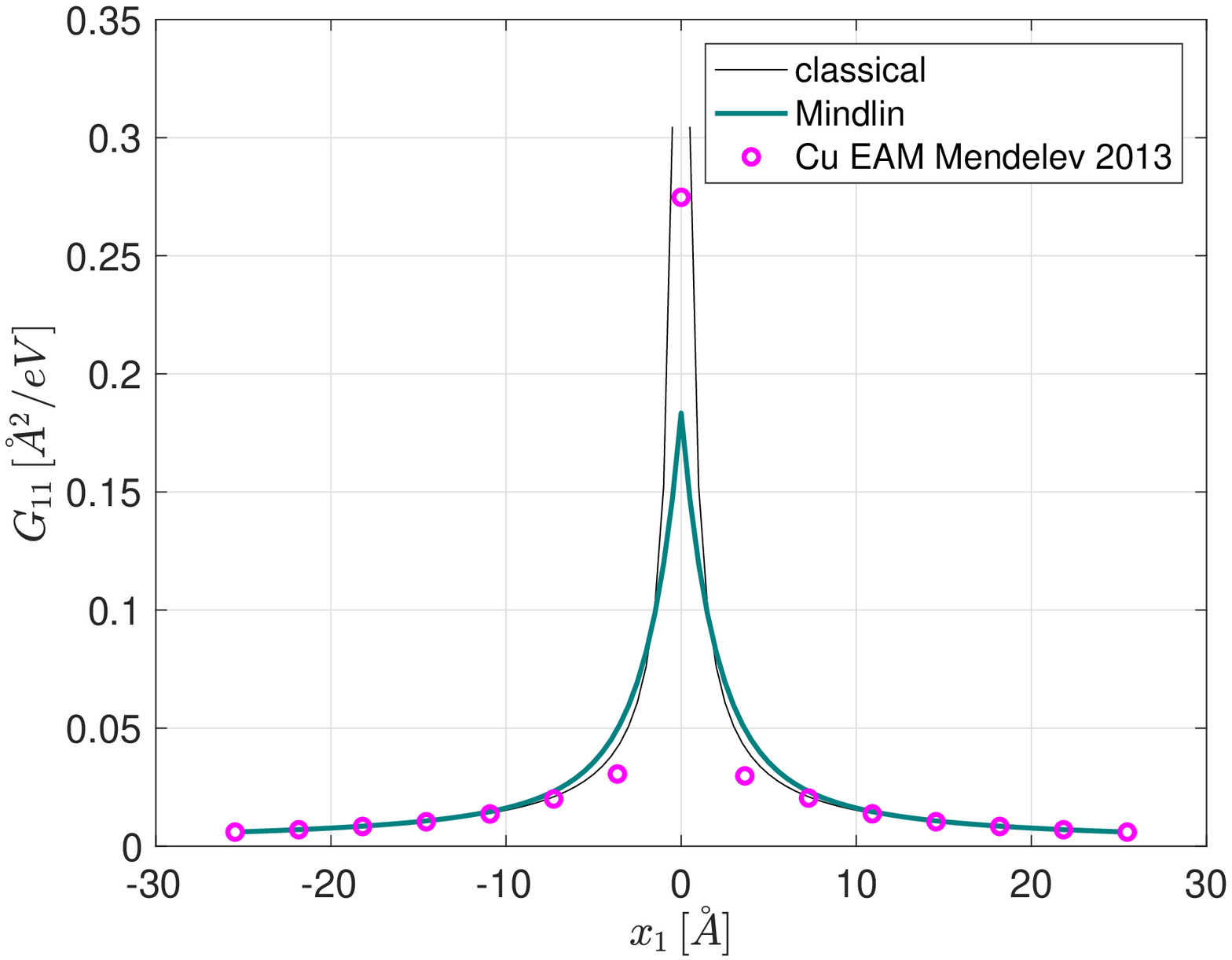}
\label{Cu_fcc_EAM_Mendelev_2013_G_11_x1}
}
\subfloat[Cu EAM]{
\includegraphics[width=0.45\textwidth]{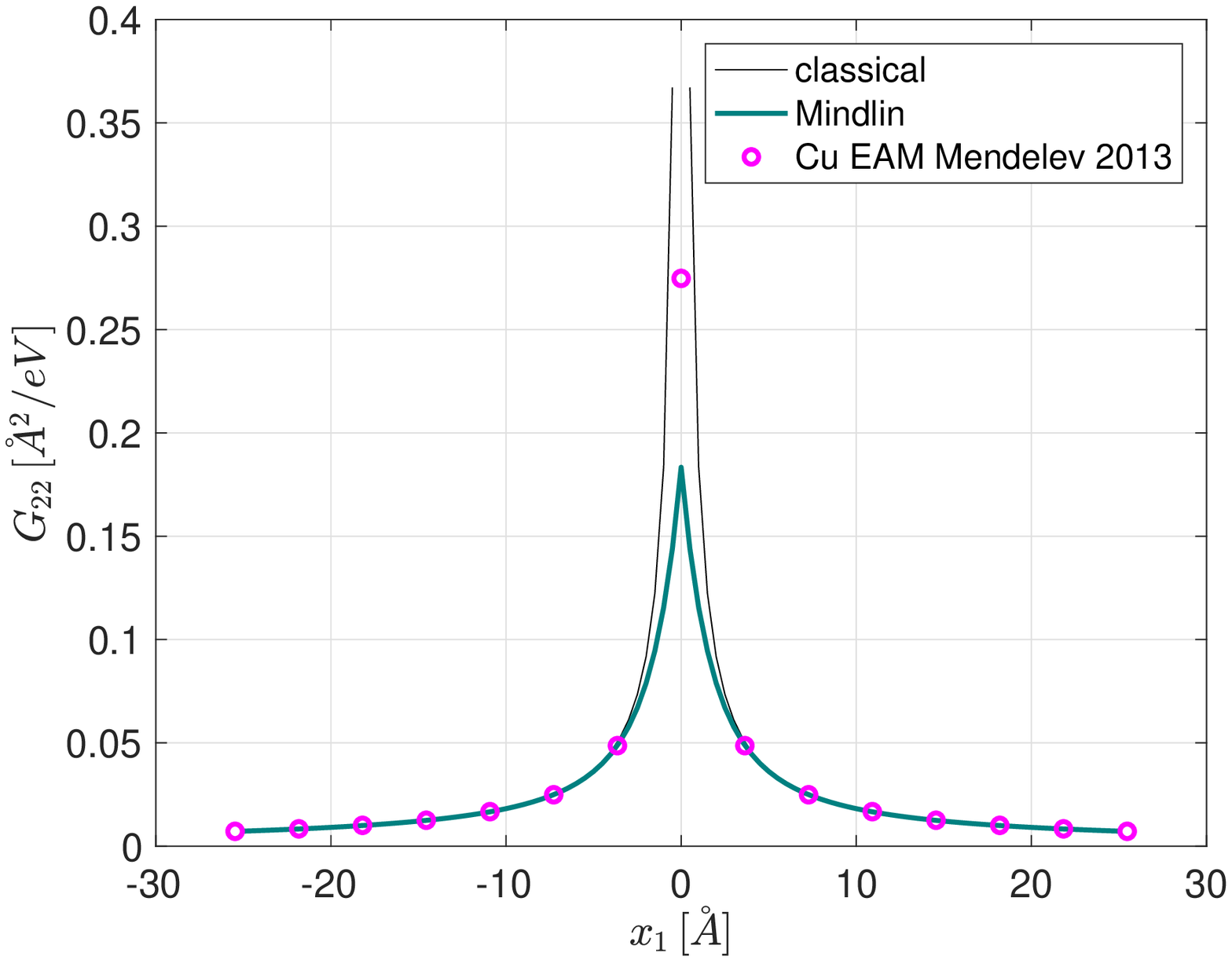}
\label{Cu_fcc_EAM_Mendelev_2013_G_22_x1}
}\hfill
\subfloat[Cu MEAM]{
\includegraphics[width=0.45\textwidth]{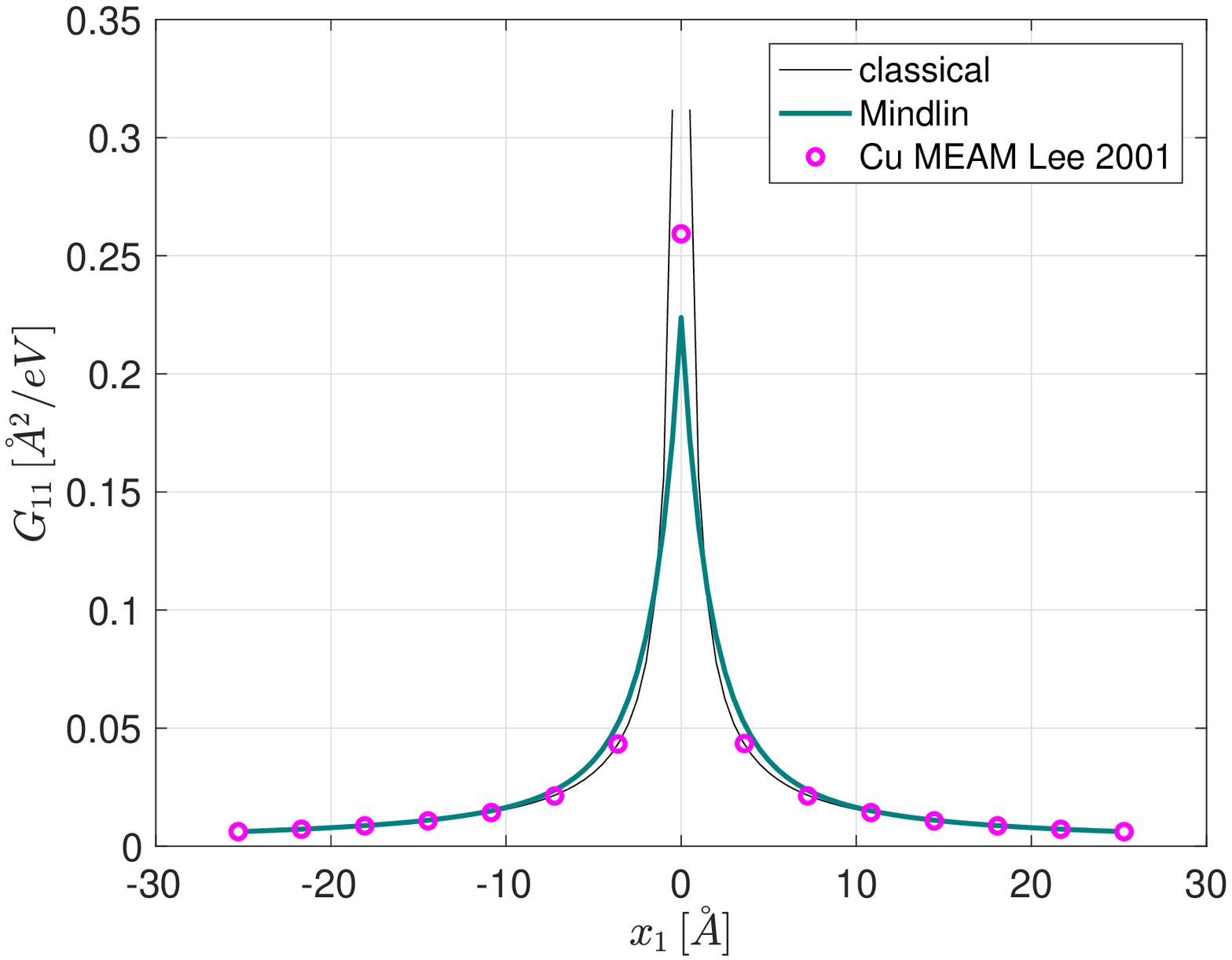}
\label{Cu_fcc_MEAM_Lee_2001_G_11_x1}
}
\subfloat[Cu MEAM]{
\includegraphics[width=0.45\textwidth]{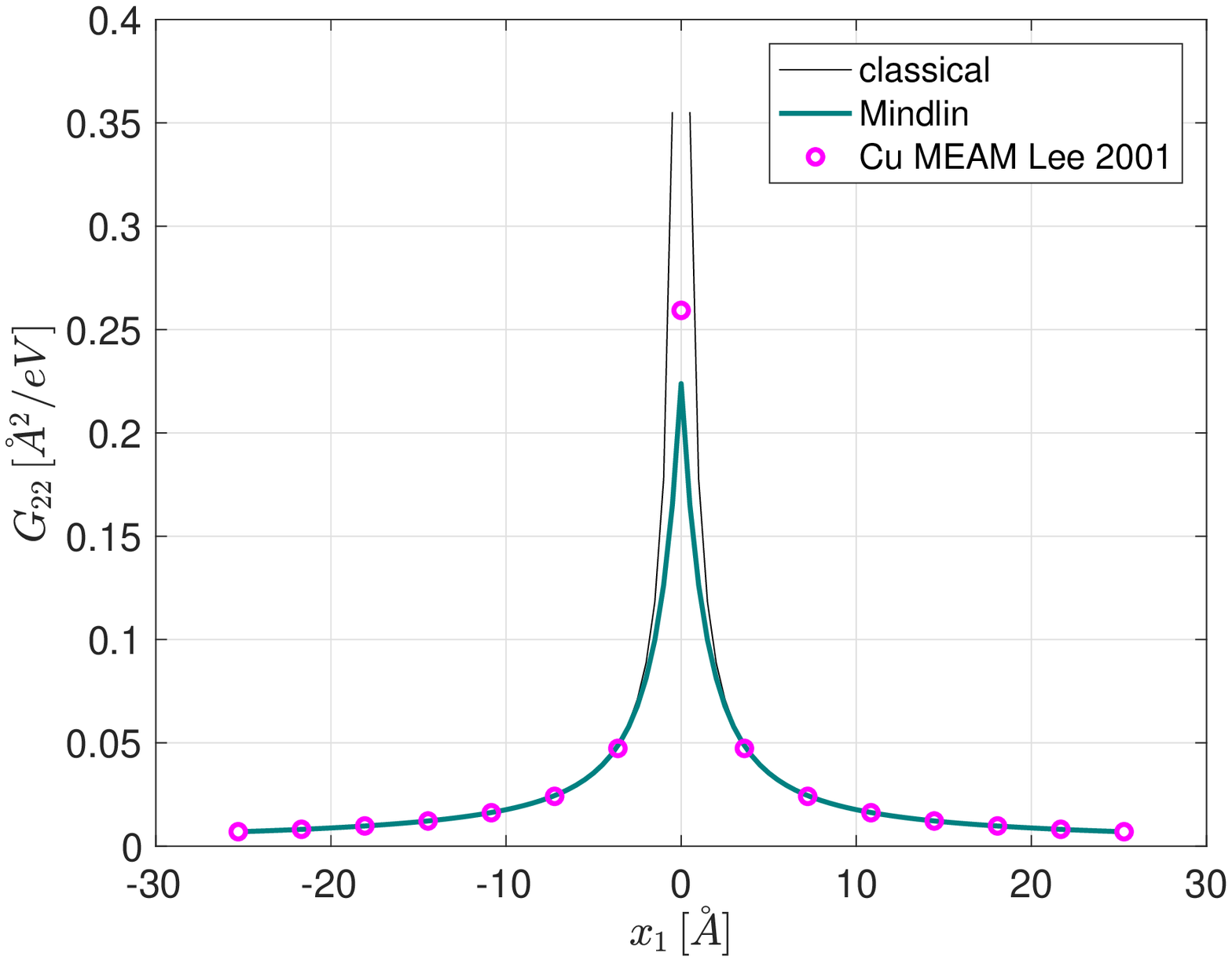}
\label{Cu_fcc_MEAM_Lee_2001_G_22_x1}
}\hfill
\subfloat[Al MEAM]{
\includegraphics[width=0.45\textwidth]{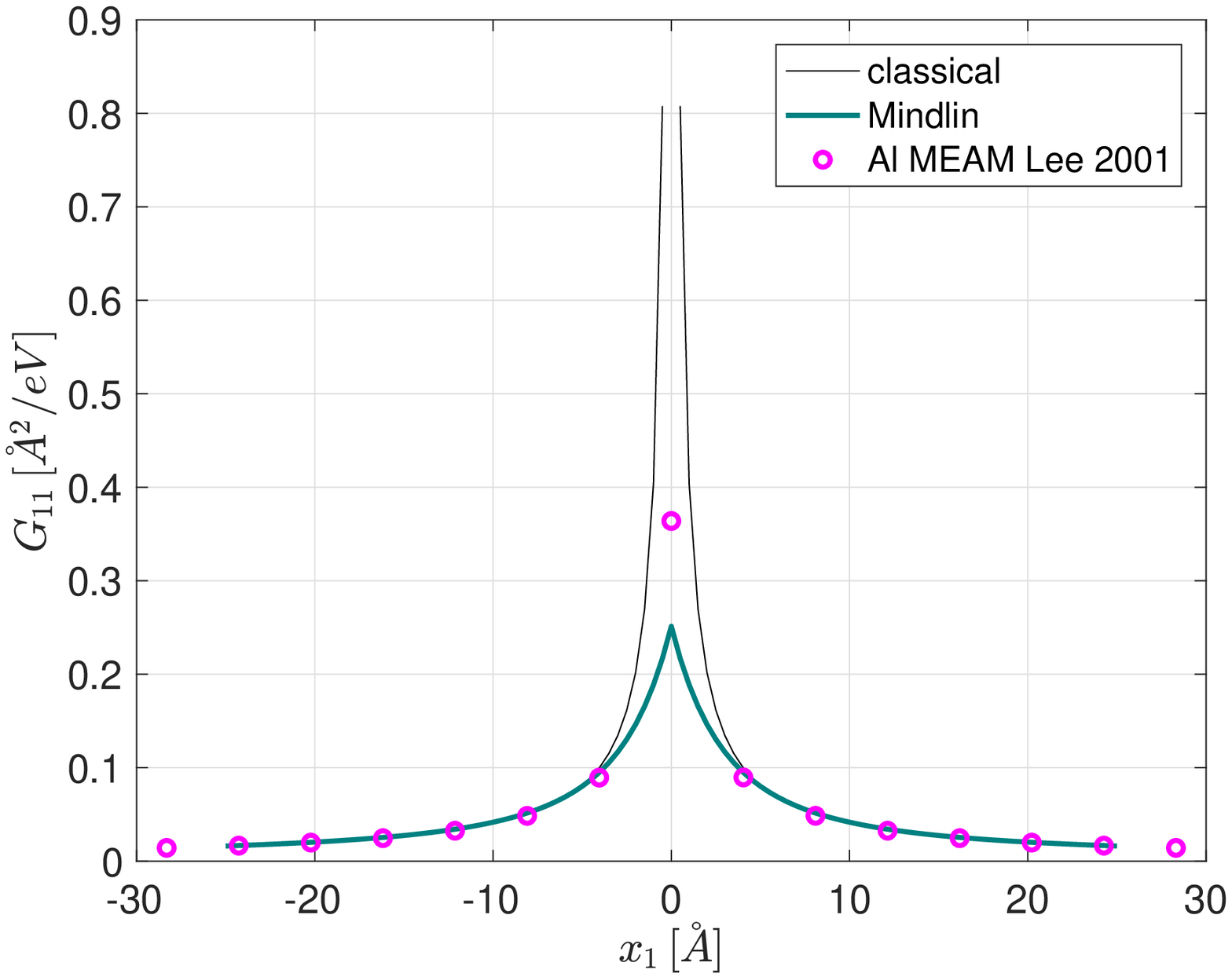}
\label{Al_fcc_MEAM_Lee_2001_G_11_x1}
}
\subfloat[Al MEAM]{
\includegraphics[width=0.45\textwidth]{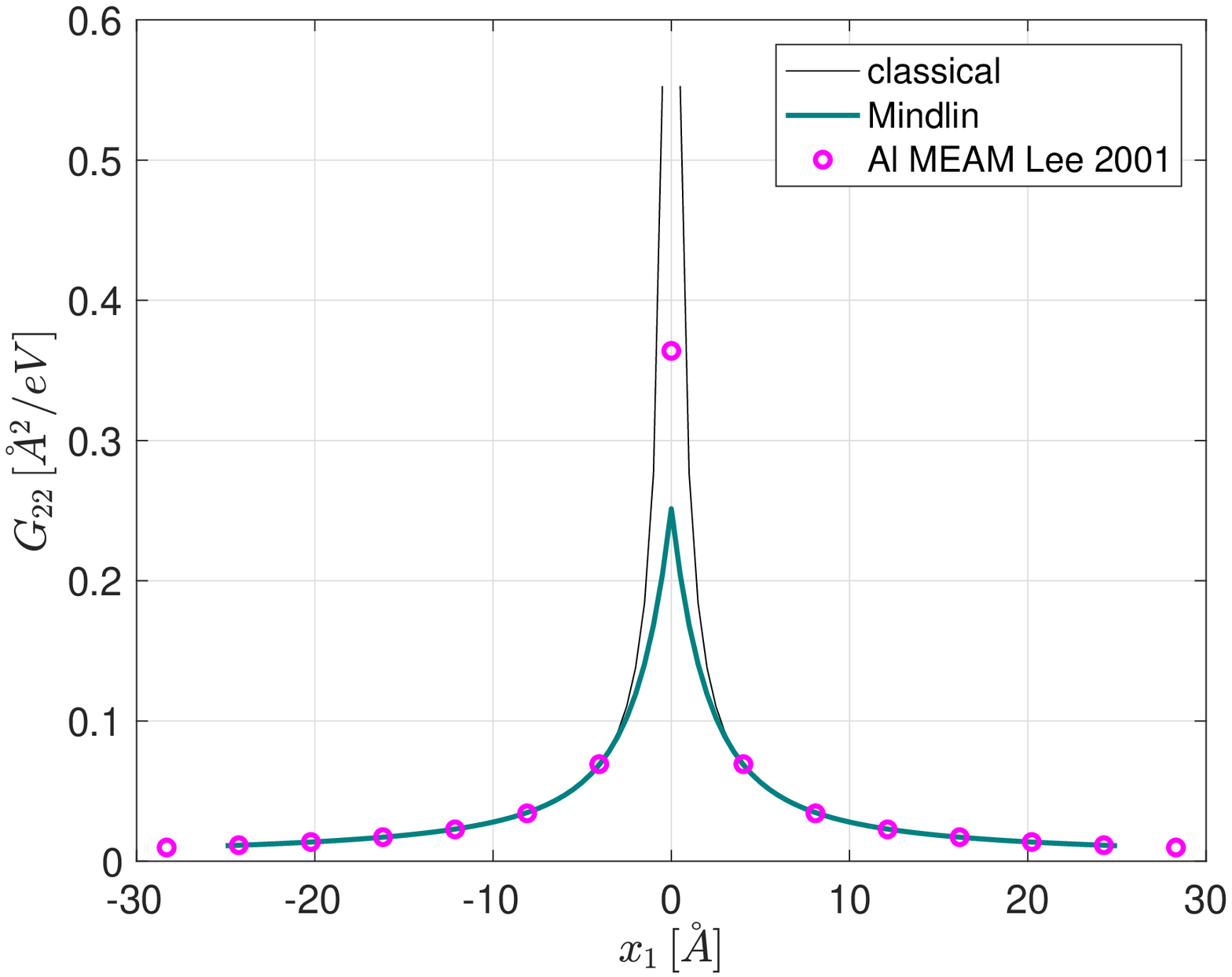}
\label{Al_fcc_MEAM_Lee_2001_G_22_x1}
}\hfill
\caption{Components of the Green tensor for  fcc Al and Cu, and comparison to atomistic calculations obtained from the interatomic potentials \cite{kimlee2001} and \cite{kimmendelev2008}. 
\protect\subref{Cu_fcc_EAM_Mendelev_2013_G_11_x1}-\protect\subref{Cu_fcc_EAM_Mendelev_2013_G_22_x1} Cu for EAM potential \cite{kimmendelev2008}.
\protect\subref{Cu_fcc_MEAM_Lee_2001_G_11_x1}-\protect\subref{D_Cu_fcc_MEAM_Lee_2001} Cu for MEAM potential \cite{kimlee2001}.
\protect\subref{Al_fcc_MEAM_Lee_2001_G_11_x1}-\protect\subref{Al_fcc_MEAM_Lee_2001_G_22_x1} Al for MEAM potential \cite{kimlee2001}.
}
\label{MScomparison}
\end{figure*}

Simulation results are shown in Fig.~\ref{MScomparison}, where we compare the Green tensor components $G_{11}(x_1,0,0)$ and $G_{22}(x_1,0,0)$. Despite the fact that these potentials were never fitted to gradient-elastic constants, it can be observed that the analytical predictions are in good agreement with  MS calculations, with a maximum error at the origin in the order of 5-30\%, depending on the potential used. It should be noted that, compared to the EAM potential,  the  MEAM potential better compares to the analytical results, possibly as a result of artifacts in gradient-elastic constants  evaluated by  EAM potentials \cite{Admal16}.

\section{Summary and Conclusions\label{conclusions}}

In this paper we have derived an expression for the Green tensor of Mindlin's anisotropic strain gradient elasticity, which possesses up to  21 elastic constants and 171 gradient elastic constants in the general case of triclinic media. The Green tensor is found in terms of a matrix  kernel  integrated over the unit sphere in Fourier space. Such representation is similar to that of the classical anisotropic Green tensor, which  requires integration over the equatorial plane of the unit sphere. In contrast to its classical counterpart, however, the Green tensor of Mindlin's anisotropic strain gradient elasticity is non-singular at the origin, while its gradient is finite but discontinuous at the origin. It is shown that the non-singular Green tensor converges to the classical tensor a few characteristic lengths away from the origin. Therefore, the  Green  tensor of Mindlin's  first strain gradient elasticity can be regarded as a physical regularization of the classical anisotropic Green tensor. Moreover, existing expressions of the Green tensor found in the literature are recovered as special cases. Because the Green tensor regularizes its classical counterpart without unphysical  singularities, it offers a more realistic description of near-core elastic fields of defects in  micro-mechanics, and it provides more accurate boundary conditions for atomistic and \textit{ab-initio} energy-minimization calculations. As an illustrative example, we have computed the displacement field induced by a concentrated force acting at the origin (Kelvin problem), and compared the analytical predictions to atomistic calculations when the elastic and gradient-elastic moduli are  consistently derived from the interatomic potentials. Despite the fact that these potentials were not fitted to gradient-elastic constants, it is shown that the analytical predictions are in good agreement with  MS calculations, with a maximum error at the origin in the order of 5-30\%, depending on the potential used.

\small{

\section*{List of Abbreviations}

PDE: partial differential equation.
SPD: symmetric positive definite.
KIM: Open Knowledgebase of Interatomic Models.
API:  application programming interface.
EAM: embedded atom method.
MEAM: modified embedded atom method.

\section*{Declarations}

\noindent\textbf{Availability of data and materials.}
Elastic and gradient-elastic material constants used to obtain the results in section \ref{Kelvin} are freely available as part of the Open Knowledgebase of Interatomic Models (KIM).
\medskip

\noindent\textbf{Competing Interest.}
The authors declare that they have no competing interests.
\medskip

\noindent\textbf{Funding.}
G.P.  acknowledges the support of the U.S. Department of Energy, Office of Fusion Energy, through the DOE award number DE-SC0018410, the Air Force Office of Scientific Research (AFOSR), through award number FA9550-16-1-0444, and the National Science Foundation, Division of Civil, Mechanical and Manufacturing Innovation (CMMI), through award number 1563427 with UCLA.
N.A. acknowledges the support of  the US Department of Energy's Office of Fusion Energy Sciences, Grant No. DE-SC0012774:0001.
M.L. gratefully acknowledges a grant from the 
Deutsche Forschungsgemeinschaft 
(Grant No. La1974/4-1).
\medskip

\noindent\textbf{Authors Contribution.}
G.P. and M.L. obtained the expression of the Green Tensor. N.A. and G.P. carried out the numerical analysis. All authors read and approved the final manuscript.

}


\bibliographystyle{spbasic}      


\appendix


\section{Direct derivation of Mindlin's isotropic strain gradient elasticity of form~II}
\label{DirectDerivationGTI}

Plugging \eqref{CFiso} and \eqref{DFiso} into \eqref{LN-FT12} we have
\begin{align}
 \bm G(\bm k)
 &=\frac{\left[\left(\lambda+2\mu\right)\left(1+k^2\ell_1^2\right)\bm\kappa\otimes\bm\kappa
 +\mu\left(1+k^2\ell_2^2\right)\left(\bm I-\bm\kappa\otimes\bm\kappa\right)
 \right]^{-1}}{k^2}\, .
 \end{align}
 Owing to its special structure (see footnote \ref{footNoteSpecialStructure}), the matrix in the numerator can be easily inverted. In index notation the result is
\begin{align}
G_{ij}(\bm k)
 &=\frac{\kappa_i\kappa_j}{\left(\lambda+2\mu\right)k^2\left(1+k^2\ell_1^2\right)}
 +
 \frac{\delta_{ij}-\kappa_i\kappa_j
 }{\mu k^2\left(1+k^2\ell_1^2\right)}\nonumber\\
  &=\frac{k_ik_j}{\left(\lambda+2\mu\right)k^4\left(1+k^2\ell_1^2\right)}
 +
 \frac{k^2\delta_{ij}-k_ik_j
 }{\mu k^4\left(1+k^2\ell_1^2\right)}\, .
\end{align}
Using the the general form of the Fourier transform of the derivative, the Green tensor in real space is obtained as
\begin{align}
G_{ij}(\bm x)&=-\frac{\partial_i\partial_j}{\lambda+2\mu}\mathcal{F}^{-1}\left[\frac{1}{k^4\left(1+k^2\ell_1^2\right)}\right]
-\frac{\delta_{ij}\Delta-\partial_i\partial_j}{\mu}\mathcal{F}^{-1}\left[\frac{1}{k^4\left(1+k^2\ell_1^2\right)}\right]
\label{GTisoF}\, .
\end{align}
Now consider the identity
\begin{align}
\mathcal{F}^{-1}\left[\frac{1}{k^4\left(1+k^2\ell^2\right)}\right]
&=\mathcal{F}^{-1}\left[\frac{1}{k^4}-\frac{\ell^2}{k^2}+\frac{\ell^4}{1+k^2\ell_1^2}\right]
=-\frac{1}{8\pi}\left(x+\frac{2\ell^2}{x} -\frac{2\ell^2}{x}e^{-x/\ell}\right)\nonumber\\
&=-\frac{1}{8\pi}A(x,\ell)\, .
\label{invAfunc}
\end{align}
Using \eqref{invAfunc} in \eqref{GTisoF}, the Green tensor \eqref{GT-iso} is readily recovered.

\newpage
\listoffigures
\end{document}